\documentclass[english,aps,prx,twocolumn,superscriptaddress]{revtex4-2}
\usepackage{amssymb}
\usepackage{amsfonts}
\usepackage{subfigure}
\usepackage{graphicx}
\usepackage{amsmath}
\usepackage{xcolor}
\preprint{}

\begin{document}

\title{Selective Kondo screening and strange metallicity in sliding Dirac semimetals}

\author{Hanting Zhong}
\email{These authors contributed equally to this work.}
\affiliation{School of Physics, Hangzhou Normal University, Hangzhou 310036, China}
\affiliation{School of Physics, Zhejiang University, Hangzhou 310027, China}

\author{Shuxiang Yang}
\email{These authors contributed equally to this work.}
\affiliation{Zhejiang Lab, Hangzhou 311121, China}

\author{Chao Cao}
\affiliation{School of Physics, Zhejiang University, Hangzhou 310027, China}
\affiliation{Center for Correlated Matter, Zhejiang University, Hangzhou 310027, China}
\affiliation{Institute for Advanced Study in Physics, Zhejiang University, Hangzhou 310058, China}

\author{Xiao-Yong Feng}
\email{fxyong@hznu.edu.cn}
\affiliation{School of Physics, Hangzhou Normal University, Hangzhou 310036, China}

\author{Jianhui Dai}
\email{daijh@hznu.edu.cn}
\affiliation{School of Physics, Hangzhou Normal University, Hangzhou 310036, China}
\affiliation{Institute for Advanced Study in Physics, Zhejiang University, Hangzhou 310058, China}


\begin{abstract}

Kondo screening of local moments in normal metals typically leads to hybridized conduction and valence bands separated by a Kondo gap, resulting in an insulating state at half-band filling. We show a dramatic change of this scenario in a Dirac-semimetal-based correlated system --- a bilayer honeycomb lattice heterostructure where a local moment lattice is stacked on a Dirac semimetal breaking the inversion symmetry. This system is modeled by an extended Anderson honeycomb lattice involving the real-space dependence of major interlayer hybridization parameters on the relative sliding distance along the armchair direction. First, we unveil multiple Kondo scales and successive Kondo breakdown transitions in this correlated heterostructure under sliding. Second, we demonstrate the existence of a genuine selective Kondo screening phase which is stabilized  near the A-B stack pattern and is accessible by applying interlayer voltage. Third, we find a nearly flat hybridized band located concomitantly within the Kondo gap, resulting in an unprecedented  metallic state at half-band filling. This unconventional heavy fermion state is characterized by violation of Luttinger theorem and appearance of a Van Hove singularity at the Fermi energy. The general sliding-driven band structure landscape and the implications of our results for the broad context of multiorbital Kondo physics are briefly discussed.

\end{abstract}

 \maketitle

\section{Introduction}

The heavy fermion (HF) physics driven by Kondo effect is among the most intriguing quantum phenomena in the strongly correlated electron systems ranging from conventional $f$ electron alloys to synthetic quantum structures \cite{StewartRMP1,StewartRMP2,Coleman2007,SiScience}. In the simplest situation, Kondo screening (KS)---the coherence of the single-ion Kondo effect--- develops with the formation of an entangled singlet state composed of periodic arrays of local moments and metallic bath \cite{HewsonBook}. Such situation can be captured by the Kondo lattice model (KLM) or the periodic Anderson lattice model (ALM)\cite{TsunetsuguRMP}, with a local Kondo coupling or inter-orbital hybridization as the driving force. As long as the Kondo coupling is non-vanishing, the electronic band structure of this model system is reconstructed, resulting in hybridized conduction and valence bands separated by the Kondo gap. This in turn leads to a Kondo insulator (KI) or HF metal at or away from the half-band filling, respectively.

However, while various Kondo couplings exist ubiquitously in realistic $f$-electron-active materials, whether or how KS actually develops remains puzzling. Indeed, KS is sensitive to material's band structures and variable interactions. The variations of these microscopic causes could conspiringly lead to the breakdown of KS.  In the present work, we will propose a theoretically lucid and experimentally controllable mechanism of the selective Kondo screening (SKS) driven by a partial breaking of crystalline symmetries. Before this, we should explain two additional motivations of this work.

Theoretically, it is worthwhile to recall the Kondo breakdown transition driven by strong magnetic fluctuations \cite{ColemanJPCM,SiNature,SenthilPRL,LohneysenRMP}. This transition is accompanied by a transformation of Fermi surfaces within or at the boundary of the magnetic phase in HF materials, enriching the global phase diagram \cite{DoniachPhysica,SiPhysica,ColemanJLTP}. In the terminology of Mott physics, it can be interpreted as an orbital/band-selective Mott transition driven by variable parameters such as bandwidth, occupation energy, or Coulomb interaction in the multiorbital Hubbard model \cite{PepinPRL,MediciPRL,VojtaJLTP}. More generally, the Kondo breakdown or dehybridization of $f$ electrons can take place independent of magnetic fluctuations, such as when the density of states at the Fermi level is depleted. This is in analogy to the single-ion Kondo problem in a pseudo-gap or graphene-like metallic bath where a finite Kondo coupling larger than a threshold value is required for the occurrence of Kondo effect \cite{WithoffPRL,HentschelPRB,SenguptaPRB}. In contrast to these situations, the SKS addressed here occurs when two degenerate $f$ orbitals (or sublattices) hybridize to a Dirac semimetal (DSM) bath. Its characteristic feature is the existence of a region where only one of $f$ orbitals is driven to the Kondo phase, while outside this region the $f$ orbitals are both in the decoupled or coupled phases, respectively.

Experimentally, there is a particularly suitable platform to investigate this new phenomenon, namely, the correlated bilayer systems such as the transition metal-dichalcogenide heterostructures  1T-TaSe$_2$/1H-TaSe$_2$ and MoTe$_2$/WSe$_2$ \cite{VanoNature,Ruan,ZhaoScience,ChenPRB,DalalPRR,Guerci}. By applying the electric field and gate voltages, one monolayer could be tuned to the Mott insulating phase with localized electrons while another remains metallic with itinerant electrons. Both electron density and interlayer Kondo coupling (or orbital hybridization) can be smoothly tuned in a controllable manner \cite{KennesNP,MakNN,LiNature}. The similar heterostructure with triangular lattices was previously realized in $^3$He films \cite{GreywallPRB,SiqueiraPRL} where some experimental features including the density-driven Mottness and the effective mass enhancement \cite{CaseyPRL,NeumannScience} were interpreted as due to an orbital-selective Mott transition or a Kondo breakdown \cite{NeumannScience,BenlagraPRL,BeachPRB}. Another experimental platform is the densely $f$ electron intercalated graphene bilayers.  This $f$-electron heterostructure can be realized using the molecular beam epitaxy technique, and some characteristic Kondo lattice features including the lower Hubbard and $4f$ quasiparticle flat bands below or around the Fermi level have been evidenced recently \cite{Wu}.

\begin{figure}[ht]
  \includegraphics [width=8cm]{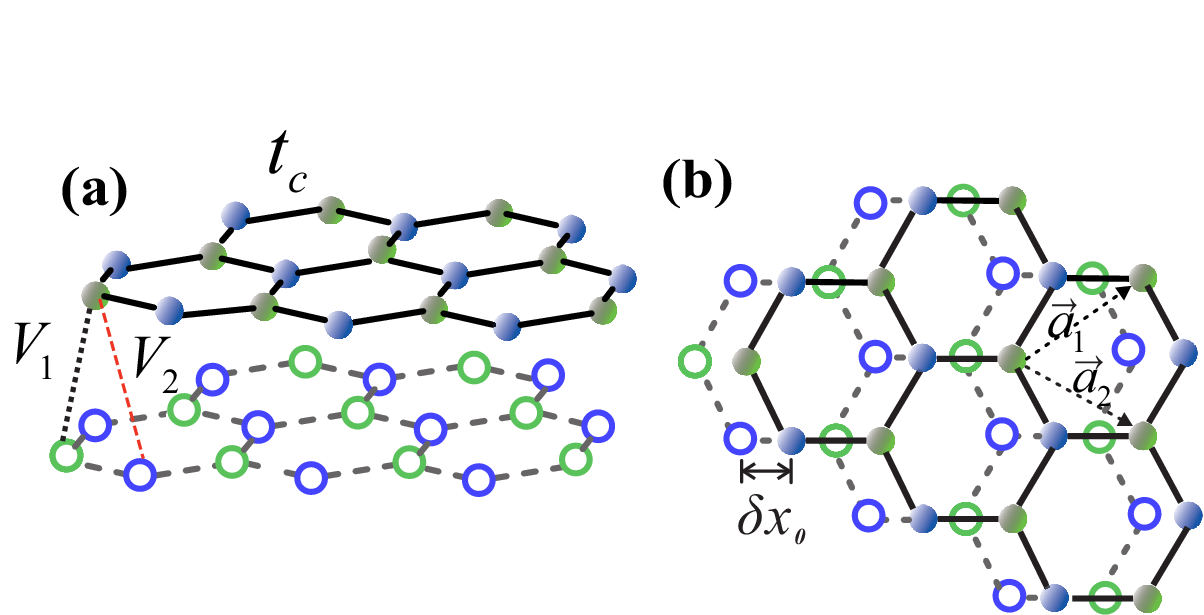}
  \caption{The stacking configuration of the heterobilayer structure. (a) Side view: The upper and lower layers are the $c$ and $f$ honeycomb monolayers, respectively, with the sublattices being distinguished by the blue and green circles.  $t_c$ denotes the $c$-electron intralayer nearest-neighbor hopping. $V_1$ and $V_2$ denote two representative interlayer hybridizations.  (b) Top view: $\delta x_0$ denotes the sliding shift from the A-A pattern along the $x$-axis. ${\vec a}_1$ and ${\vec a}_2$ are the lattice vectors.}  \label{stacking configuration}
\end{figure}

Motivated by these advances, we here consider a heterostructure composed of a localized $f$-electron honeycomb lattice stacked by the itinerant $c$-electron graphene. We assume that the $f$-monolayer is fixed, while the $c$-monolayer can slide along a given direction, chosen as the armchair direction connecting the high and low symmetry configurations \footnote{Similar to bilayer graphene, the high symmetry stack configuration is the so-called A-A pattern with the ${\cal C}_6$ symmetry where the atoms of one layer are positioned perfectly above the atoms of the other layer, the low symmetry configuration is the A-B pattern with ${\cal C}_3$ symmetry,  obtained by sliding one layer along the armchair direction at the position where half of the atoms in each layers are at hollow sites. Usually, the low symmetry configuration is energetically more stable so that the sliding process from the A-A to A-B  patterns is energetically favorable. The materials-dependent changes of the adhesion potential and the interlayer distance in the sliding process are not taken into account in the present study. They play an essential role in the structural stability  \cite{PopovPRB,MostaaniPRL},  but not in the Kondo physics}\cite{PopovPRB,MostaaniPRL}. The generic stacking configuration is shown in
Fig. \ref{stacking configuration}. Recall that in the homobilayer graphene the sliding process does not dramatically change the band structure\cite{NetoRMP,McCannRPP}, while twisting the bilayer would result in the occurrence of flat bands and rich correlated quantum phases\cite{BistritzerPNAS,CaoNature,AndreiNM}. So far the effect of the sliding process has been intensively investigated in connection with the stacking-engineered ferroelectricity in the two-dimensional van der Waals materials \cite{Li_Wu_Nano17,vdW_sliding_Science21,bilayer_boron_nitride_Science21,SternNatPhys}, while its interplay with the strong electron correlation is less explored. In the following, we shall find that the sliding process (without twisting) in the present correlated heterostructure could tune the selectivity of the multiorbital KS, the latter is delicately sensitive to the inversion symmetry breaking. The resultant unconventional flat hybridizing band adds a new ingredient to the broad context of Kondo physics.

\section{Model Hamiltonian}

We start from the model Hamiltonian which consists of three parts:
\begin{eqnarray}
\hat{\cal H} = \hat{\cal H}_c + \hat{\cal H}_f + \hat{\cal H}_{cf}.
\end{eqnarray}
The first part describes the conduction ($c$) monolayer:
\begin{eqnarray}
\hat{\cal H}_c = -\sum_{\langle\bf{i}\bf{j}\rangle\sigma}[t^{(c)}_{\bf{i}\bf{j}}\hat{c}^{\dag}_{\vec r_{\bf{i}}\sigma}\hat{c}_{\vec r_{\bf{j}}\sigma}+H.c.],
\end{eqnarray}
with ${\hat c}_{\vec r_{\mathbf{i}}\sigma}$ being the annihilation operator of conduction electrons, $\sigma=\uparrow, \downarrow$ the spin degrees of freedom, and $t^{(c)}_{\bf{i}\bf{j}}$ the hopping matrix. The subscript $\bf i$ or $\bf j$ labels the sites on the honeycomb lattice, $\vec r_{\bf i}$ is the position vector in this monolayer. Reproducing the essential features of the Dirac semimetal bath allows to assume the nonzero intralayer hopping parameter $t^{(c)}_{\bf{i}\bf{j}}=t_c>0$ for the nearest neighbor sites only. Similarly, the second part describes the $f$-monolayer:
\begin{eqnarray}
\hat{\cal H}_{f} &=& -\sum_{\langle\bf{i}\bf{j}\rangle\sigma}[t^{(f)}_{\bf{i}\bf{j}}\hat{f}^{\dag}_{\vec R_{\bf{i}}\sigma}\hat{f}_{\vec R_{\mathbf{j}}\sigma}+H.c.] \nonumber\\
&+& E_0\sum_{\bf{i}\sigma}\hat{n}_{f\bf{i}\sigma} + U\sum_{\bf{i}}\hat{n}_{f\bf{i}\uparrow}\hat{n}_{f\bf{i}\downarrow},
\end{eqnarray}
with $\hat{n}_{f\bf{i}\sigma}=\hat{f}^{\dag}_{\vec R_{\mathbf{i}}\sigma}\hat{f}_{\vec R_{\mathbf{i}}\sigma}$  and $\vec R_{\bf i}$ the corresponding position vector in this monolayer. Here, a nearest neighbor hopping energy $t^{(f)}_{\bf{i}\bf{j}}=t_{f}$ for $f$ electrons is introduced without losing the generality. In the realistic situation, $|t_f/t_c|$ is very small and we shall take $t_{f}=0$ in most calculations, together with the occupation energy $E_0<0$ and the on-site Coulomb interaction $U\rightarrow \infty$. Namely, this monolayer is assumed in the Mott phase at half filling in the absence of hybridization. The third part is the interlayer hybridization term:
\begin{eqnarray}
\hat{\cal H}_{cf} =\sum_{\{\bf{i}\bf{j}\}\sigma}[ V_{\bf{ij}} \hat{c}^{\dag}_{\vec r_{\bf{i}}\sigma}\hat{f}_{\vec R_{\bf{j}}\sigma}+ H.c.]
\end{eqnarray}
with  $V_{\bf{ij}}$ being the generic interlayer hybridization matrix elements.

There are two favorable stacking configurations as in the bilayer graphene\cite{NetoRMP,McCannRPP}: the A-A and A-B (or Bernal) patterns with the ${\cal C}_6$ and ${\cal C}_3$ symmetries, respectively. The sliding process considered here smoothly interpolates these patterns with the reflection symmetry ${\cal M}_x$.
In each monolayer the location of the $\bf j$-th site can be assigned by $\bf{j}=({\bf n},\eta)$, with ${\bf n}=(n_1,n_2)$ being a pair of integers labelling the unit cells and $\eta=A, B$ the even or odd sublattices, respectively. The $f$-layer is fixed in the $z=0$ plane, so $\vec R_{\bf{j}}=n_1\vec a_1+n_2\vec a_2+\frac{a_0}{2}\epsilon_{\eta} \vec e_x$, with $\vec a_1=a_0(\frac{3}{2} \vec e_x+ \frac{\sqrt 3}{2}\vec e_y)$, $\vec a_2=a_0(\frac{3}{2} \vec e_x- \frac{\sqrt 3}{2}\vec e_y)$ being the two lattice vectors in the base plane ($a_0$ the distance between the nearest neighboring sites), $\epsilon_{\eta}=-1$ or $1$ for $\eta=A$ or $B$, respectively, and $\vec e_{x}$, $\vec e_{y}$, $\vec e_{z}$ the respective spatial unit vectors.
Accordingly, in the $c$-layer, $\vec r_{\mathbf{i}}=m_1\vec a_1+m_2\vec a_2+\frac{a_0}{2}\epsilon_{\eta} \vec e_x+a_z \vec e_z+\delta x_0\vec e_x$, with $a_z$ being the interlayer distance and $\delta x_0$ the relative shift along the $x$-axis, the armchair direction. The cases $\delta x_0=0$ and $\delta x_0=a_0$ correspond to the A-A and A-B patterns, respectively. Further increasing $\delta x_0$ will result in a cyclic evolution A-A$\rightarrow$ A-B$\rightarrow$ M $\rightarrow$ B-A$\rightarrow$ A-A with a period $3a_0$, where the M pattern corresponds to the middle point $\delta x_0=3a_0/2$. Therefore, it is sufficient to consider the sliding distance $0\leq \delta x_0\leq 3a_0/2$. The inversion symmetry ( the simultaneous exchange between the odd and even sublattices) is preserved in the A-A and M patterns only.

The hybridization matrix elements decrease rapidly with the geometric distance $d({\bf i},{\bf j})=|\vec r_{\bf{i}}-\vec R_{\bf{j}}|$, assumedly following the rough behavior $V_{{\bf ij}}=V[\frac{a_z}{d({\bf i},{\bf j})}]^{\zeta} e^{-|d({\bf i},{\bf j})-a_z|/\xi}$, with $\zeta \geq 0$ being the materials-dependent exponent, $\xi\sim a_0$ the characteristic decay length, and $V$ the single interlayer hybridization strength. Hence, it is legitimate to consider the major elements coming from the intra-cell and the nearby (the nearest or the next nearest ) inter-cell hybridizations, denoted by $V^{(\Delta_1,\Delta_2)}_{\eta\eta'}$, with $\eta,\eta'=A, B$ and $\Delta_i\equiv m_i-n_i=0,\pm 1$. This approximation corresponds to $\xi/a_0=1\sim {\sqrt 3}$. Without losing the generality,
We calculate the band structure landscape with various $\delta x_0$ by simply assuming $V_{{\bf ij}}=V[\frac{a_z}{d({\bf i},{\bf j})}]^2$ (corresponding to $\zeta=2$) and using $\xi=\sqrt 3 a_0$ as a planar-distance cut-off (see more explicit numerical schemes in Appendix A).

In order to elucidate the essential features of the KS phases, we will then isolate the ideal situation with a single hybridization parameter, i.e., the strongest interlayer hybridization. This parameter can be identified as the nearest-neighbor hybridization $V_1=V_{\langle \bf{ij}\rangle} (=V)$  in the A-A and A-B patterns ($V^{(0,0)}_{\eta\eta}$ and $V^{(0,0)}_{AB}$), respectively.  It plays the most crucial role in the formation of the SKS. To investigate the stability of this phase, we will further include the next-nearest-neighbor hybridization $V_2=V_{\langle\langle \bf{ij}\rangle\rangle} =\alpha V$ as a perturbation, with $\alpha$ being a small tuning parameter.

The Hamiltonian at the limit $\delta x_0=0$, the A-A pattern, was previously studied (in the case $t_f=0$ and $V_2=0$) by using various methods, including the slave-boson technique \cite{FengPRL,FengDKSM}. In this situation, the hybridization matrix respects the inversion symmetry. With finite $V_1(=V)$, the two $c$ bands start to hybridize the doubly degenerated local $f$ orbital, resulting in the four Bloch bands provided $V_1$ is larger than a critical value $V_c$, above which the KS occurs. As explicitly shown in the Appendices $B$ and $C$, $V_c$ slowly decreases by turning on $V_2$, without essential change in the band structure. The Dirac semimetal feature is reflected by a pair of hybridised Dirac-like conduction and valence bands that are symmetrically separated by the Kondo gap. So, the system is in the KI phase at half-filling.

The present study focuses on the sliding process (for $0\leq\delta x_0\leq 3a_0/2$) and pays special attention to the point $\delta x_0=a_0$, corresponding to the A-B pattern. This situation is fairly non-trivial since the ${\cal C}_3$ symmetry is restored although the inversion symmetry is lost due to interlayer hybridization. The previous band structure result must be altered and the existence of the Kondo phase is in question.

\section{Method}

In order to clarify the possible Kondo phase in the present problem, we consider the limit $U\rightarrow \infty$ and solve the model by using the slave boson mean-field method \cite{ColemanPRB,NewsAP}. The $f$ electron operator is then represented by a fermionic operator $\hat d_{\textbf j \sigma}$ and a bosonic operator $\hat b_{\bf j}$ such that  $\hat f_{\vec R_{\textbf j}\sigma}^{\dag}=\hat d_{\bf{j}\sigma}^{\dag}\hat b_{\bf{j}}$. The large $U$ limit imposes the no-double occupation constraint $\hat b^{\dag}_{\bf{j}}\hat b_{\bf{j}}+\sum_{\sigma}\hat d^{\dag}_{\bf{j}\sigma}\hat d_{\bf{j}\sigma}=1$ at each lattice sites.
The hybridization matrix induces the effective elements $\tilde V_{\bf{ij}}=\langle \hat b_{\bf j}\rangle V_{\bf{ij}}$, where $\langle \hat b_{\bf j}\rangle$ is the expectation value of the slave boson on a given state and can be always treated as real. The constraint is implemented by introducing a Lagrange multiplier $\lambda_{\bf j}$ in the path integral approach (see Appendix E). In this approach, the annihilation or creation operators are represented by the respective field variables, $(c_{{\textbf j \sigma}}, d_{{\textbf j \sigma}}, b_{{\textbf j}})$ or $(\bar c_{{\textbf j \sigma}}, \bar d_{{\textbf j \sigma}}, b^{*}_{{\textbf j}})$, with the implied temperature-dependence.

Due to the inversion symmetry breaking, the expectation value $\langle \hat b_{\bf j}\rangle$  at the site $\bf j=(\bf n,\eta)$ on the groundstate is $\eta$-dependent, while it is uniform in $\textbf n$ due to the lattice translational invariance. Therefore, we can introduce two real order parameters, $r_{\eta}=\langle \hat b_{{\bf n},\eta}\rangle$, for each sublattices $\eta=A, B$. The solution with $r_{\eta}>0$ implies the occurrence of KS in $\eta$-sublattice. Then the substitution $b_{\eta}\rightarrow r_{\eta}$ could be invoked in the mean-field approach. Similarly, the Lagrange field is also $\eta$-dependent and is represented by the parameter $\lambda_{\eta}$ accordingly.
All these mean-field parameters should be determined self-consistently.

With the above considerations, the effective mean-field Hamiltonian is expressed in the momentum space as
\begin{eqnarray}
\mathcal{H}_{MF} = \sum_{\mathbf{k}\sigma}\bar\Psi_{\mathbf{k}\sigma}H_{\mathbf{k}\sigma}\Psi_{\mathbf{k}\sigma}+E_C
\end{eqnarray}
with $\Psi_{\mathbf{k}\sigma}=(c_{A\mathbf{k}\sigma},c_{B\mathbf{k}\sigma},d_{A\mathbf{k}\sigma},d_{B\mathbf{k}\sigma})^{T}$ being the corresponding Fourier transformed fields and the momentum $\textbf k$ in the hexagonal Brillouin zone (BZ). Here, $E_C=L\sum_{\eta}\lambda_{\eta}(r_{\eta}^2-1)$, $L$ is the total number of unit cells, and the Hamiltonian matrix reads
\begin{eqnarray}
H_{\bf{k}\sigma}=\left(
                       \begin{array}{cccc}
                         0 & \varepsilon_{c,\bf k}& r_A h_{AA} & r_B h_{AB} \\
                        \varepsilon^*_{c,\bf k}& 0 & r_A h_{BA} & r_B h_{BB} \\
                         r_A h^*_{AA} & r_A h^*_{BA} & E_0+\lambda_A & \varepsilon_{f,\bf k}\\
                         r_B h^*_{AB} & r_B h^*_{BB} & \varepsilon^*_{f,\bf k} & E_0+\lambda_B \\
                       \end{array}
                     \right)
\end{eqnarray}
with $\varepsilon_{c,\bf k}=-t_c f_{\bf k}$, $\varepsilon_{f,\bf k}=-t_f r_A r_B f_{\bf k}$,
$f_{\bf k}=e^{ik_x}[1+e^{-i{\bf k}\cdot \vec a_1}+e^{-i{\bf k}\cdot \vec a_2}]$, and $h_{\eta\eta'}=\sum_{\Delta_1\Delta_2}V^{(\Delta_1,\Delta_2)}_{\eta\eta'}e^{i[{\bf k}\cdot (\vec a_1\Delta_1+\vec a_2\Delta_2)]}$ for $\Delta_i=0,\pm 1$.

The solution of the eigenvalue problem of $H_{\bf{k}\sigma}$ involves four ( spin-degenerate ) quasiparticle bands $E_{{\bf k}\sigma m}$ ($m=1,2,3,4$) in the Kondo phase, see details in Appendix B.
The free energy in the thermal equilibrium at temperature $T$ is given by
\begin{eqnarray}
F=-\frac{1}{\beta}\sum_{\mathbf{k}\sigma m}\ln{[1+e^{-\beta (E_{{\bf k}\sigma m}-\mu)}]}+E_C,
\end{eqnarray}
where, $\beta=1/(k_B T)$, $\mu$ is the chemical potential determined by the total electron number
$\sum_{{\bf k}\sigma m}n_F(E_{{\bf k}\sigma m}-\mu )=N$, with $n_F(x)=(e^{\beta x}+1)^{-1}$ the Fermi function.
Minimizing the free energy with respect to $r_{\eta}$ and $\lambda_{\eta}$ leads to a set of mean-field equations
\begin{eqnarray}
\sum_{{\bf k}\sigma m} n_F(E_{{\bf k}\sigma m}-\mu )\frac{\partial E_{{\bf k}\sigma m}}{\partial\lambda_{\eta}}+L(r^2_{\eta}-1)=0,\\
\sum_{{\bf k}\sigma m} n_F(E_{{\bf k}\sigma m}-\mu )\frac{\partial E_{{\bf k}\sigma m}}{\partial r_{\eta}}+2Lr_{\eta}\lambda_{\eta}=0.
\end{eqnarray}

We solve these equations in the low temperature regime approaching the limit $\beta\rightarrow\infty$ as for the ground state at the half-band filling ($N=4L$). In calculations, we first adopt the previously mentioned relationship $V^{(\Delta_1,\Delta_2)}_{\eta\eta'}= V[a_z/d^{(\Delta_1,\Delta_2)}_{\eta\eta'}]^2$ and tune the model parameter $V$ for several choices of $\delta x_0$, with  $d^{(\Delta_1,\Delta_2)}_{\eta\eta'}=\sqrt{[\frac{3}{2}(\Delta_1+\Delta_2)+\frac{1}{2}(\epsilon_{\eta}-\epsilon_{\eta'})+\delta x_0]^2+\frac{3}{4}(\Delta_1-\Delta_2)^2+a^2_z}$. Other model parameters are fixed at $a_z=1.5 a_0$, $t_c=1$, $t_f=0$, $E_0=-5$. The inverse temperature is fixed at $\beta=400$ which is sufficient for identifying the ground state by convergent numerical calculations. Next, we focus on the A-B pattern using the simplified $V_1$-$V_2$ hybridizations, by tuning the model parameters $V$, $\alpha$, and $t_f$, respectively. Our primary strategy is to search for the region of $V$ where the nonzero solutions for $(r_A, r_B)$ are obtained at the zero temperature limit. In this region, the corresponding Kondo energy scales are identified as the finite temperatures $T_{K,\eta}$ beyond which $r_{\eta}$'s are suppressed.

\section{Main results}

\subsection{Evolution of the Kondo phase}

The previous mean-field equations for $(r_{\eta},\lambda_{\eta},\mu)$ are solved numerically in a wide range of model parameters, especially for variable interlayer hybridization parameters. These numerical solutions depend delicately on the sliding distance $\delta x_0$.  Due to the ${\cal M}_x$ symmetry and using the length cut-off (chosen as $\xi=\sqrt 3 a_0$), there are $5\times 4=20$ major hybridization elements $V^{(\Delta_1,\Delta_2)}_{\eta\eta'}$ as illustrated in Appendix A.1. All these hybridization elements are functions of the hybridization strength $V$, chosen as the nearest-neighbor interlayer hybridization. Different cut-off schemes are compared, see Appendix A.2, showing a roughly similar $\delta x_0$-dependence. Therefore we here present the results obtained by using a representative cut-off scheme with $\xi=\sqrt{3}a_0, \zeta=2$ and $a_z=1.5 a_0$.

First, we find that the numerical solutions for non-vanishing $r_A$ or $r_B$ do not exist if $V$ just slightly increases from zero. This fact implies that the Kondo phase could not be accessed by treating $V$ as a perturbation.  Then, we search for the solutions of $(r_{A}, r_{B})$ as functions of the shift $\delta x_0$ from the large $V$ side, i.e., $V=5,4,3$, respectively. A relatively larger $V$ necessitates the occurrence of KS for the sake of gapless nature of the Dirac metal bath\cite{FengPRL}. Further, we determine the critical hybridization strength $V=V_c$ where $r_A=0$ or $r_B=0$ by decreasing $V$ from this side as far as possible. By this way we are able to draw the general phase diagram which exhibits the evolution of the Kondo phase in terms ($\delta x_0$,$V$).

As shown in the upper panel of Fig. \ref{dx_dependence} (see Fig. \ref{shift} in Appendix C for more details), the KS occurs at $\delta x_0=0$ with finite $r_A=r_B$ for $V=3,4,5$, implying that the KS takes place in both $f$ electron sublattices with the same Kondo energy scale, owing to the inversion symmetry. This phase is enhanced by increasing $V$.  When $\delta x_0>0$, $r_A$ becomes smaller than $r_B$, indicating the emergence of two distinct Kondo scales due to lack of the inversion symmetry. Moreover, both $r_A$ and $r_B$ decrease when $\delta x_0/a_0$ goes through $0.5$ where they show rather distinct features: $r_B$ is about $0.18, 0.40$ and $0.55$ at $\delta x_0/a_0=0.5$ as the local minima for $V=3,4$ and $5$, respectively; while $r_A$ is about $0.1,0.2$ and $0.3$ at that point for $V=3,4$ and $5$, respectively. Moreover, $r_A$ decreases further and approaches zero for further increased $\delta x_0 $. Specifically, $r_A\sim 0$ at $\delta x_0/a_0=0.6$ even for $V=5$, After that, while $r_A$ persists at zero, $r_B$ increases again with a maxima at $\delta x_0/a_0=1$. In order to find the boundary of the Kondo phase, we have also solved $r_A$ and $r_B$ by further decreasing the hybridization strength ($V<3$). For each value of $\delta x_0$, we have extracted the critical $V_{c,A}$ for $r_A=0$ (dashed blue line ) or $V_{c,B}$ for $r_B=0$ (solid orange line) as shown in the lower panel in Fig. \ref{dx_dependence}, respectively.

These results indicate three facts: (1) For a given sliding shift $\delta x_0$, there is a  critical hybridizations $V_{c,\eta}$ for $r_{\eta}$=0 below which the KS for $f$ electrons in the $\eta$-sublattice does not exist. For all values of $\delta x_0$, we have finite $V_{c,B}\leq V_{c,A}$ for $\delta x_0$ in between $(0,3a_0/2)$ (or $V_{c,B}\geq V_{c,A}$ for $\delta x_0$ in between $(3a_0/2,3a_0)$). In particular, $V_{c,A}\sim V_{c,B}$ when $\delta x_0\rightarrow 0$ or $\delta x_0\rightarrow 3a_0/2$, although the corresponding $r_A$ and $r_B$ are still distinct except in the vicinity of the A-A or M-patterns recovering the inversion symmetry.    (2) There is a region (in $\delta x_0=0.5a_0\sim a_0$) where $f$ electrons are hybridized in the B sublattice ($r_B>0$) but dehybridized in the A sublattice ($r_A=0$). This is the genuine orbital-( or lattice-) SKS with a single non-vanishing Kondo energy scale. (3) While various hybridization parameters are variably at play in the intermediate region, the nearest-neighbor hybridization $V_1$  crucially influences the formation of the KS in the region around the A-B pattern.

Therefore, there are three distinct groundstate phases: the fully decoupled local moment phase (FLM) where $V<V_{c,A}$ and $V<V_{c,B}$; the fully coupled KS phase (FKS) where  $V>V_{c,A}$ and $V>V_{c,B}$; the genuine SKS phase (SKS) where $V_{c,B}< V< V_{c,A}$. These phases are clearly distinguished in the lower panel of Fig. \ref{dx_dependence}.
Notice that the values of $V_{c,A}$ and $V_{c,B}$ depend on the microscopic model parameters and the cut-off scheme delicately. Hence the dependence of these values on the sliding distance as well as the boundaries of these phases in realistic materials such as MoTe2/WSe2 or $f$-doped graphene requires more detailed first-principle band structure calculations. However, the existence of these phases is robust in a wide range of the model parameters as evidenced in  Appendix C. The influences of different $f$-electron levels $E_0$ on the values of $V_{c,\eta}$ are also shown in Appendix C.

\begin{figure}[ht]
  \includegraphics [width=8cm]{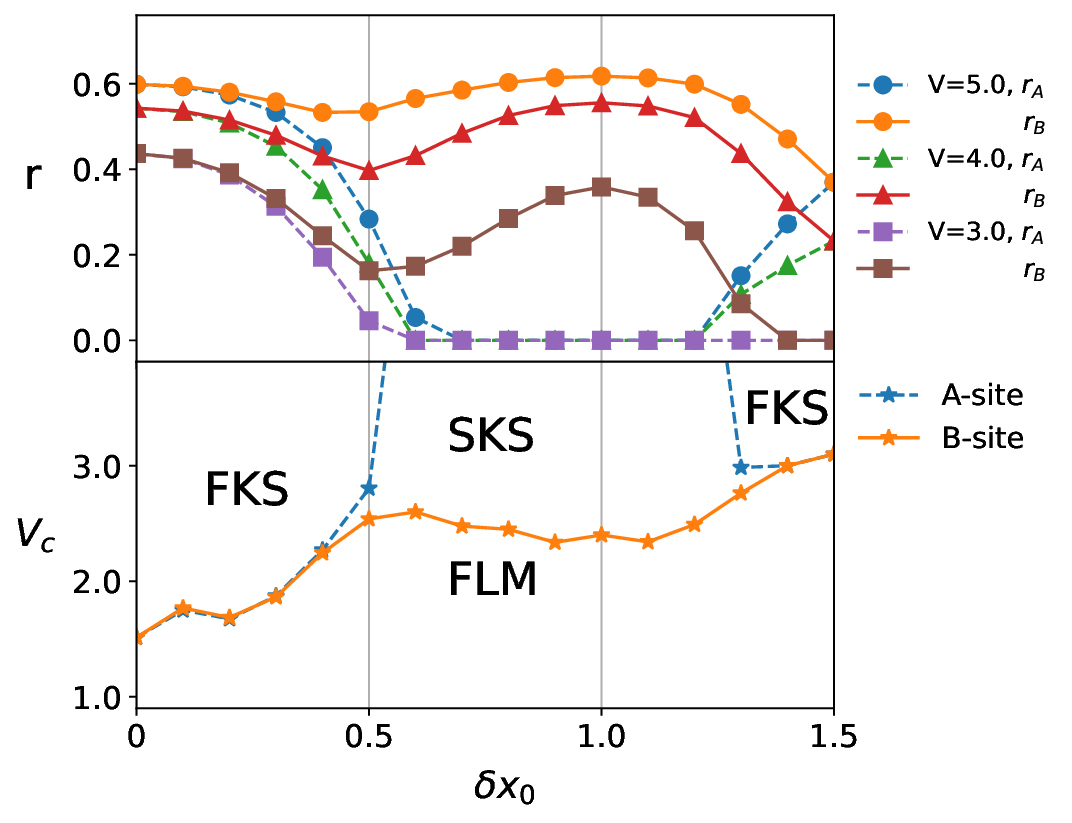}
  \caption{$\delta x_{0}$-dependence of $r_{\eta}$ for different $V$'s (upper panel)
and the extracted critical $V_{c,\eta}$'s (lower panel) with $\beta=400$ and $E_{0}=-5$. Here the cut-off scheme with $a_z=1.5, \xi=\sqrt{3}, \zeta=2$ is used. In the lower panel there are three regions separated by $V_{c,A}$ (dashed blue line) and $V_{c,B}$ (solid orange line), representing the fully decoupled local moment phase(FLM), fully coupled Kondo screened phase(FKS), and the genuine selective Kondo screened phase(SKS), respectively. Note that the shift has a period of $3a_0$ and is symmetric around $\delta x_{0}=1.5$ ($a_0$ is taken to be unit).}
  \label{dx_dependence}
\end{figure}

\subsection{Existence and stability of SKS}

Previous results hint the optimized SKS ($r_A=0$ and $r_B$ takes a maxima) in the A-B pattern ($\delta x_0=a_0$) where the ${\cal C}_3$ symmetry is recovered. In this situation $V_{c,A}$ is much larger than $V_{c,B}$, so we can examine this phase more pertinently around the critical hybridization $V_c\equiv V_{c,B}$ with the most dominative interlayer hybridization parameter $V_1=V$ only. In order to clarify the stability of the SKS, we now focus on this pattern starting from the limiting situation with $t_f=0$ and $\alpha=0$, i.e., the $f$ electrons being exactly local and only subject to the nearest neighbor interlayer hybridization. In this ideal situation, the $f$ orbital from the $A$-sublattice is apparently decoupled so that the original four-band model Hamiltonian is reduced to an interacting three-band model involving two $c$ orbitals and one $f$ orbital from the $B$-sublattice only.  Such situation is the $B$-sublattice extension of the conventional two-channel single-ion Kondo impurity problem where the overscreening of a Kondo impurity may lead to the non-Fermi liquid behavior\cite{NozieresJPhys,SenguptaPRB}. Hence it is of particular interesting as to what extend the $B$-sublattice $f$ electrons can be screened by the semimetallic bath. It is also necessary to examine the instability of this phase under the influences of small $t_f$ and $\alpha$. These two issues are clarified below numerically within the four-band Hamiltonian.

\begin{figure}[ht]
\hspace*{-0.3cm}
\centering\includegraphics [scale=0.42]{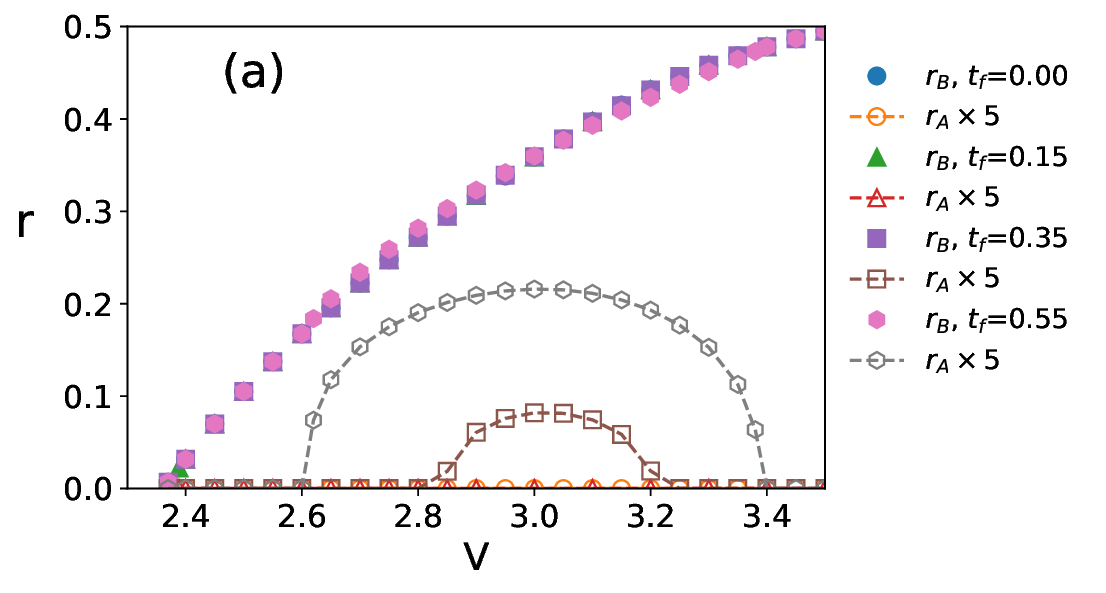}
\centering\includegraphics [scale=0.42]{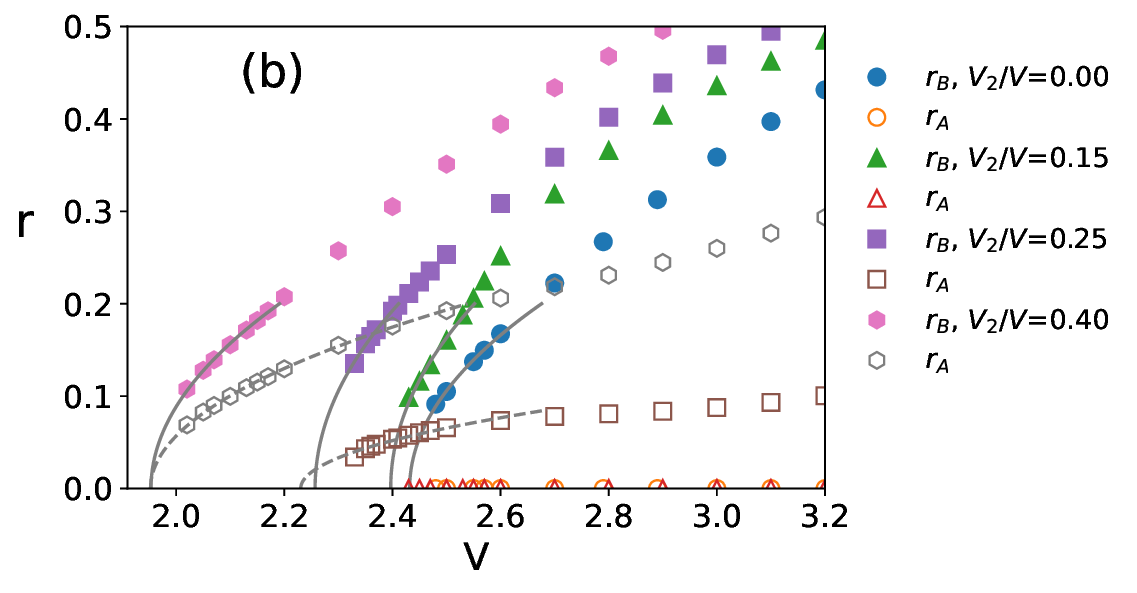}
  \caption{Mean-field solutions of $r_A$ (open symbols ) and $r_B$ (filled symbols) in the A-B pattern with $t_c=1$, $E_0=-5$, $\beta=400$ using the simplified ($V_1,V_2$) hybridization parameters. (a) Twith fixed $V_2=0$ (upper panel (a)) or fixed $t_f=0$ (lower panel (b)). (a) $V_2$ is fixed at zero, while $t_f=0, 0.15, 0.35$, and $0.55$. The solution of $r_B$ depends very weakly on $t_f$ and is nonzero when $V$ is above a critical value $\sim 2.3$. Nonzero $r_A$ exists only in a narrow intermediate region around $V\sim 3.0$. This region increases slightly with $t_f$. Note that the $r_A$ is multiplied by a factor of 5 for better illustration. (b) $t_f$ is fixed at zero, while $\alpha=V_2/V=0, 0.15, 0.25$, and $0.4$. The solid and dash lines are quadratic fitting from the last several data points.}
  \label{AB solution}
\end{figure}

\begin{figure}[ht]
  \includegraphics  [scale=0.5]{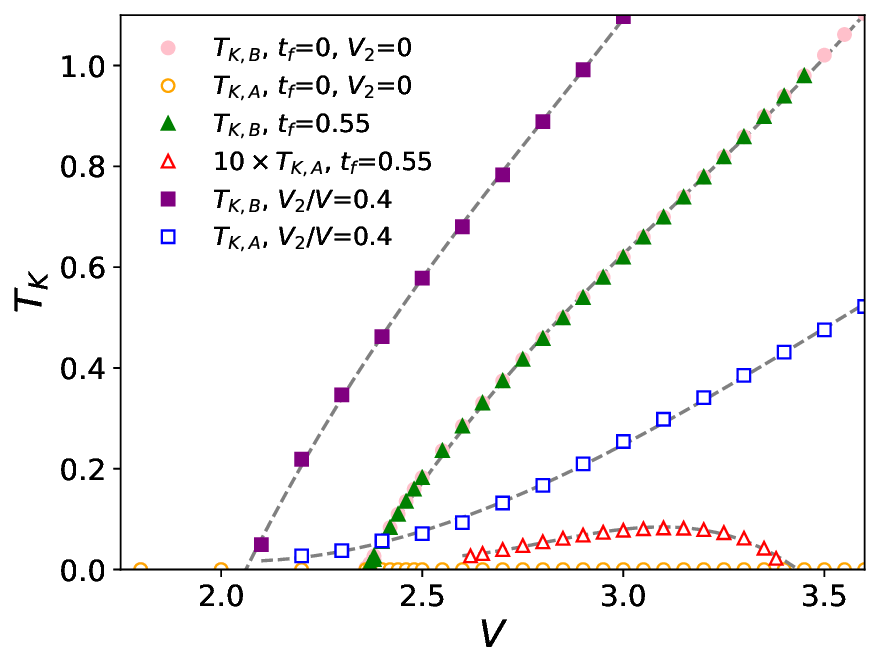}
  \caption{Hybridization strength dependence of the Kondo temperatures for B-sublattice (filled symbols) and A-sublattice (open symbols) for $t_f=0, V_2=0$, $t_f=0.55, V_2=0$ and $t_f=0, V_2/V=0.4$. Note that $T_{K,A}$ for $t_f=0.55, V_2=0$ is multiplied by a factor of ten to make it visible. }
  \label{Tk}
\end{figure}

The numerical solutions of $V$-dependent $(r_A, r_B)$ in the four-band Hamiltonian are plotted in Fig. \ref{AB solution}.  Generally, we find a nonzero critical value around $V_c\equiv V_{c,B}=2.38$ below which $r_A=r_B=0$, implying that the $f$ electrons in both sublattices are dynamically decoupled due to the insufficient hybridization ($V_1<V_c$). This result resembles to the single-ion Kondo problem\cite{WithoffPRL,IngersentPRB,VojtaPRB,UchoaPRL}, manifesting the pseudo-gap feature of the Dirac semimetal bath.  When $V_1>V_c$ (but still much smaller than $V_{c,A}$), on the other hand, we find that $r_B$ increases while $r_A$ remains zero, indicating the emergence of a genuine SKS, i.e., KS occurs in one $f$ orbital (or sublattice) while Kondo breakdown in the other.  This scenario is in agreement with the one obtained from the three-band Hamiltonian. Consequently, there is a single non-vanishing Kondo energy scale $T_K$ (as function of $V$) associated with this peculiar multiorbital phase as plotted in Fig. \ref{Tk}. Notice that the similar result is not accessible within the perturbative approach in the corresponding single-ion Kondo impurity in graphene\cite{UchoaPRL}, manifesting the non-perturbative Kondo coherence in the present lattice problem. Based on the present numerical result and assuming negligible quantum fluctuations, a formal path integral approach to this SKS phase is briefly illustrated in Appendix E.

When $t_f$ increases from zero, we find that $V_c$  remains almost unchanged except for a narrow region $t_f\sim 0.35-0.55$ where $r_A$ grows but is significantly smaller than $r_B$. In this narrow region, the KS occurs with two distinct Kondo scales ($T_{K,A}$ and $T_{K,B}$ associated with the two respective sublattice $f$ electrons) as shown in
Fig. \ref{Tk}. When $\alpha=V_2/V$ increases from zero, $V_c$ is reduced moderately but still nonzero, showing the robustness of the Kondo breakdown transition. More details about the influence of $\alpha$ are shown in Fig.\ref{AB solution}(b). Since both $t_f$ and $\alpha$ are relatively small in realistic materials, the SKS with a critical $V_c$ is stable near the A-B pattern in a wider range of the hybridization parameters.

\subsection{Strange metallicity}

The existence of nonzero critical hybridization $V_c$ explored here manifests the non-perturbation nature of the present KS physics which cannot be accessed by smoothly tuning the hybridization parameter from the completely decoupled phase. Consequently, the metallic state emerging from the SKS phase is expected to be beyond the description of the conventional Fermi liquid. Evidence for this unconventional metallic state comes also from the Luttinger theorem which states that in a Fermi liquid its Fermi volume does not change under the adiabatic change of interaction parameters\cite{LuttingerPR,OshikawaPRL}.

In the present system, the lattice translational invariance is preserved both in the paramagnetic phase as well as in the simple collinearly ordered phases formed by the $f$ moments, since there are even number of $f$ sites in each unit cell. Now assume that we can tune the hybridization strength $V$ from zero to a moderately large value across $V_{c,B}$ and $V_{c,A}$, so that the system undergoes from the fully decoupled to the SKS and the fully screened phases, successively. In the full-decoupled phase, only $c$ electrons contribute to the Fermi volume of the original DSM, leading to the small Fermi volume ${\cal V}_{\rm {DSM}}=\frac{(2\pi)^2}{2v_0}(2n_{c} \mod 2)$. Here, the factor of 2 and $\mod 2$ account for the spin degeneracy and filled bands, respectively, and $v_0$ the unit cell volume. This is in agreement with Luttinger's theorem \cite{LuttingerPR,OshikawaPRL} although the Fermi surface closes at $n_c=1$ for the ideal DSM.
In the full-screened Kondo phase, both $c$ and $f$ electrons contribute to the formation of the HF liquid (HFL), with the total number of electrons per unit cell being given by $n_{\rm {tot}}=2n_c+2n_f=2n_c+2$. Usually, the Fermi volume should be large in this case. But here the corresponding Fermi volume is ${\cal V}_{\rm {HFL}}= \frac{(2\pi)^2}{2v_0}(n_{\rm{tot}} \mod 2)$, still in agreement with Luttinger's theorem and the system is in the HFL phase for $n_c\neq 1$ or the KI phase for $n_c=1$.

In the SKS phase, however, the Fermi volume is contributed by $c$ electrons and half of the $f$ electrons (in the B-sublattice), given by ${\cal V}_{\rm {SKS}}=\frac{(2\pi)^2}{2v_0}(2n_{c}+1 \mod 2)$. This intermediate Fermi volume is large and violates Luttinger's theorem since there is a jump in the Fermi volume either from the full-decoupled phase or from the full-screened phase.   Moreover,  the SKS phase at the half filling ( $n_c=1$ ) is necessarily in a metallic state. This is in striking contradiction to the conventional Kondo lattice system where a Kondo insulator state appears at half-filling.

\begin{figure}[ht]
\hspace*{-0.0cm}\centering\includegraphics [scale=0.35]{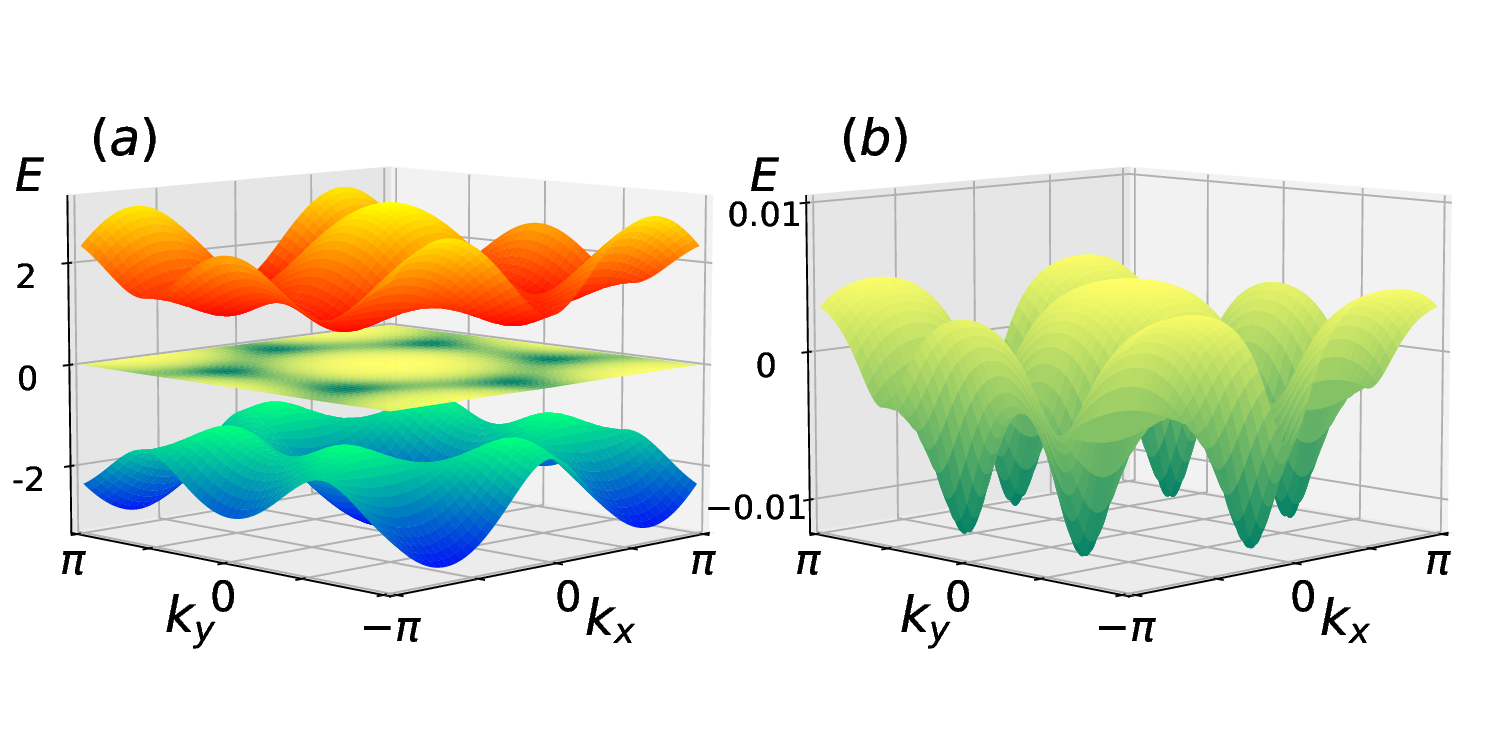}
  \caption{The two-dimensional band structure in the A-B pattern using the reduced three-band model involving the nearest neighbor interlayer hybridization parameter $V_1=V$. Left: The three hybridization bands in the SKS phase are plotted with $t_c=1$, $E_0=-5$, $V_2=0$, $t_f=0$, and $\beta=400$. A moderate $V=3$ is used for illustration. The decoupled fourth band is located at $E_0=-5$  and is not plotted here. Right: the amplified picture of the nearly flat band exhibiting dips at the Dirac points in $ \textbf k$-space). } \label{AB bandstructure}
\end{figure}

To reveal the essential features of the SKS phase, we calculate the band structure in the A-B pattern at half-filling using the reduced three-band Hamiltonian at $V=3$. There are three diagonalized energy bands for the hybridized quasiparticles as plotted in Fig. \ref{AB bandstructure}(a). The first (upper) and third (lower) bands, which resemble the conventional heavy fermion valence and conduction bands, are well-separated by a relatively larger direct band gap $\sim 2.0$ (the conventional Kondo gap).
In addition, the second (middle) band is nearly flat, located around the Fermi energy and separated from both the upper and lower bands as shown in Fig. \ref{AB bandstructure}(a). Its bandwidth is proportional to $r^2_B$ and is almost suppressed near the boundary of the Kondo phase.  Nevertheless, this band still shows the $\textbf{k}$-dependent dispersion as could be detected in the right panel in Fig. \ref{AB bandstructure}(b) (where the band structure is amplified just for visibility):  it has several dips located at the Dirac points (the $K$-point) in the $\textbf{k}$-space as a result of the ${\cal C}_3$ symmetry.

\begin{figure}[ht]
\centering\includegraphics [scale=0.55]{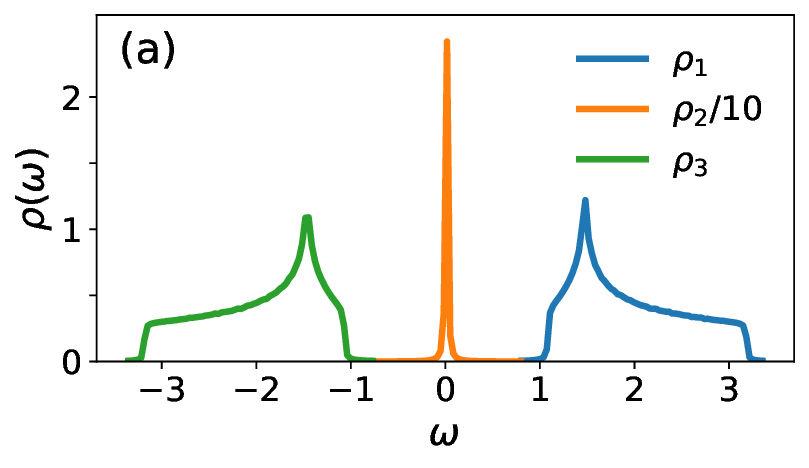}
\centering\includegraphics [scale=0.55]{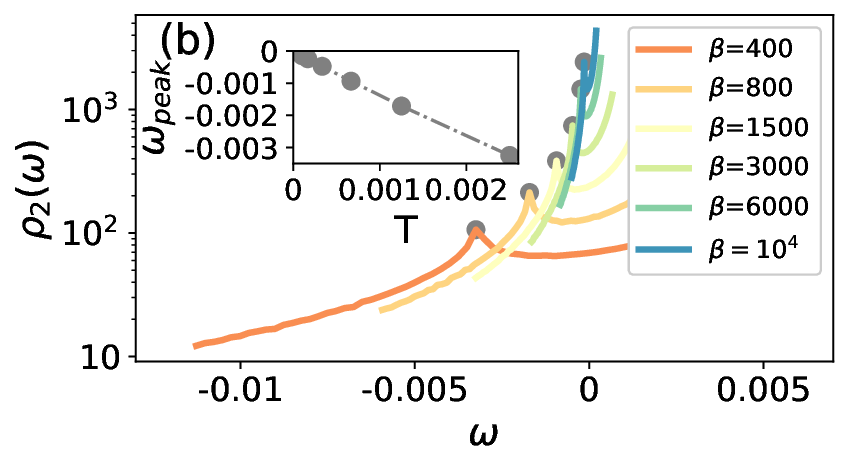}
\centering\includegraphics [scale=0.55]{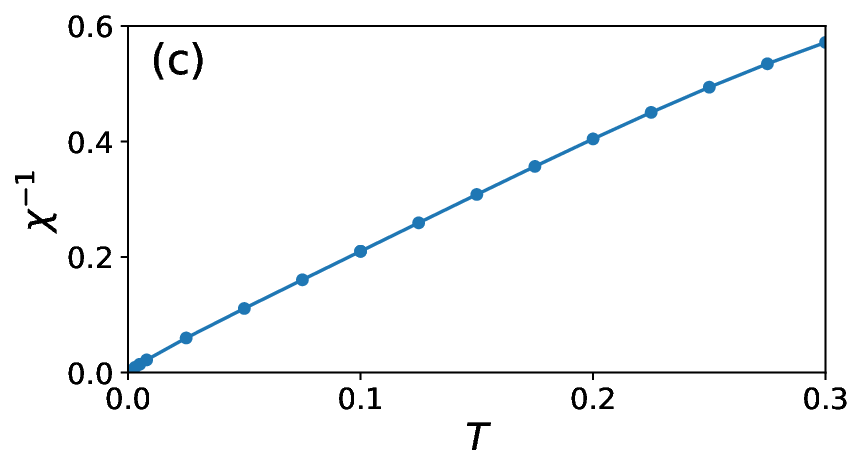}
  \caption{ (a) The density of states (DOS). Three peaks contributed from the respective bands at $\beta=400$ (upper panel) are well-separated. The DOS around the central peak is multiplied by a factor $1/10$  for illustration.  (b) A zoom-in look for the DOS of the second band with temperatures ranging from $\beta=400$ to $\beta=10000$ (middle panel). The peak position ( the gray full circle) shifts towards to the zero frequency, and its location $\omega_{peak}$ is plotted as a function of temperature in the inset.  (c) Temperature dependence of the inverse spin susceptibility $\chi^{-1}$ when approaching the low temperature limit (lower panel). Other parameters are fixed at $E_0=-5$, $t_f=0$, $V=3$, and $\alpha=0$ in the A-B stack pattern.}
  \label{DOS}
\end{figure}

The energy separation between these three hybridizing bands can be more clearly observed in the density of states (DOS) as shown in Fig. \ref{DOS}(a). A three-peak structure is readily seen due to the presence of energy gaps between them. Consequently, the contributions to each peaks can be distinguished by the respective bands. The corresponding DOS's are then denoted by $\rho_1(\omega), \rho_2(\omega)$ and $\rho_3(\omega)$, respectively. More remarkably, there is a Van Hove singularity in $\rho_2(\omega)$ contributed from the nearly flat band around the Fermi energy as shown in Fig. \ref{DOS}(a) ( Notice that $\rho_2(\omega)$ is reduced by $1/10$ in Fig. \ref{DOS}(a) for better illustration ). This singularity is clearly due to the hybridized quasiparticles near the $M$-point as shown in the inset of Fig. \ref{AB bandstructure}. In order to determine its precise location in energy, we calculated the DOS's of this band for much lower temperatures ranging from $\beta=400$ to $\beta=10000$, plotted in Fig. \ref{DOS}(b). We find that the singularity is closed to the Fermi level at the low-temperature limit as shown in the inset of
Fig. \ref{DOS}(b). As shown in Appendix D for the system deep in the SKS phase with a relatively larger hybridization ($V=5$), the peak feature is still very prominent and the temperature dependence of $\rho_2(\omega)$ is very weak so it indeed reflects the groundstate electronic behavior.

Further, we calculate the low-temperature spin susceptibility $\chi$ ( per unit cell) as shown in Fig. \ref{DOS}(c) where the inverse susceptibility $\chi^{-1}$ is plotted for the visibility. The susceptibility increases very sharply approaching the zero temperature limit due to the appearance of the Van Hove singularity. Based on the Stoner's criterion, it would lead to a possible ferromagnetic instability triggered by a very small residual interaction between the quasiparticles in realistic materials. This primary ordering tendency is consistent with the Lieb's argument \cite{lieb1989two} given the inequivalent $A$ and $B$ sublattices in the A-B pattern.

Given the violation of the Luttinger theorem and the strong Van Hove singularity in the metallic SKS phase, other correlated effects are also possible. In particular, the non-Fermi liquid behavior is expected to dominate not only in the vicinity of the critical point $V_{c,B}$, but also in the entire SKS region. One signature of such non-Fermi liquid behavior would be the small itinerant spin/charge-density wave (with materials-dependent wave vectors) with fractional excitations \cite{senthil2004weak}, another is a possible spinon Fermi surface contributed by the decoupled $A$-sublattice $f$ electrons due to quantum fluctuations or RKKY-like interactions. All these correlation effects deserve further investigations.

\section{Summary and Discussions}

In summary, unlike the twisting in the bilayer graphene, the sliding process in the present correlated heterostructure does not change the unit cell but induces rich multiorbital KS physics owing to the semimetallic nature of the conduction electron bath and the inversion symmetry breaking.  This allows a general description for the real-space interlayer hybridization matrix elements proportional to a single hybridization strength $V$, usually chosen as the nearest-neighbor interlayer hybridization parameter. The hybridization matrix elements decrease substantially with distance and the major elements can be reasonably approximated using the cutoff scheme. The systematical evolution of the Kondo phase can be mapped out by tuning the sliding distance $\delta x_0$ and the hybridization strength $V$. Three phases, i.e., the full-decoupled, the full-coupled, and the selectively coupled phases, are distinguished by two critical hybridization strengths $V_{c,A}$ and $V_{c,B}$, corresponding to the onsets of the KS for the local $f$ electrons in the A- or B-sublattice, respectively.  The corresponding Kondo scales, represented by the mean-field parameters $r_{A}$ and $r_{B}$, are distinct due to the inversion symmetry breaking ( except for $\delta x_0=0, 3a_0/2$, where the inversion symmetry is restored ). It turns out further that the main features of these phases can be captured by the cases of $\delta x_0=0$ and $\delta x_0=a_0$, where the high symmetry ${\cal C}_6$ and low symmetry ${\cal C}_3$ are respected. In these two stack patterns, the nearest-neighbor interlayer hybridization element $V_1$ ($=V$) is optimized, dominating over all other hybridization elements, while the next-nearest-neighbor interlayer hybridization element $V_2$ ($=\alpha V$) can be treated as a major perturbation.  Due to the Dirac metal nature of the host, a critical hybridization strength ($V_c>0$) is necessary for the occurrence of the full KS near the A-A pattern and the genuine SKS near the A-B pattern. Complications arise in the intermediate situation ($0<\delta_0<a_0$) where the ${\cal C}_6$ or ${\cal C}_3$ symmetry breaks down to the ${\cal M}_x$ symmetry, resulting in the delicate $V_1$-$V_2$ Kondo frustration.

Among many sliding-driven phenomena observed so far, the most remarkable observation is the metallic state in the SKS phase even at the half-filling. A moderately large hybridization strength $V_{c,B}<V<V_{c,A}$ is required, and conventional perturbative treatment with respect to $V$ is not accessible to this region where the Luttinger theorem is violated. In this sense, this metallic state is strange, distinguished from the Fermi liquid. Notice that in conventional heavy fermion systems the conduction and valence hybridization bands are separated by a finite Kondo gap, resulting in a Kondo insulator at the half-band filling.  In the present model, similar conduction and valence hybridization bands with a moderate bandwidth could become active only above the $3/4$-band fillings or below the $1/4$-band fillings, corresponding to the low carrier density limit. For comparison, the strange metallicity in the present SKS phase is due to the in-gap hybridizing metallic state at the half-band filling, characterized by a very narrow bandwidth. It is remarkable that such a nearly-flat band still accommodates a Van Hove singularity close to the Fermi energy at the half-band filling, resulting in a divergent susceptibility at the low-temperature limit.

Here we stress that the violation of the Luttinger theorem is a robust physical feature due to the pseudo-gap nature of the Dirac metal bath and the sliding-driven inversion symmetry breaking. First, the present problem represents a lattice extension of the single-impurity Kondo problem in the pseudo-gap bath and the critical hybridizations resemble to the critical Kondo coupling obtained by the numerical renormalization group calculation for the corresponding Kondo model \cite{ingersent1996behavior}. Second, the lift of the degeneracy between $V_{c,B}$ and $V_{c,A}$ is due to the inversion symmetry breaking so that the existence of the SKS phase is robust although the values of $V_{c,B}$ and $V_{c,A}$ determined here are based on the mean-field method. Third, the violation of Luttinger theorem is a rigorous consequence of the paramagnetic SKS phase since half of the $f$ electrons (in the A-sublattice ) are dynamically decoupled from the Dirac metal bath.
Therefore, our results uncover a class of new Kondo physics driven by breaking the inversion symmetry in the general multiorbital Kondo systems.
While rich correlation effects such as ferromagnetic order and unconventional superconductivity as observed in other twisted bilayer systems can be expected, we would like to briefly discuss below the sliding-driven landscape of band structure, the accessibility of the KS transition, and the theoretical implications as concluding remarks.

\begin{figure}
\hspace*{-0.5cm}\centering\includegraphics [scale=0.40]{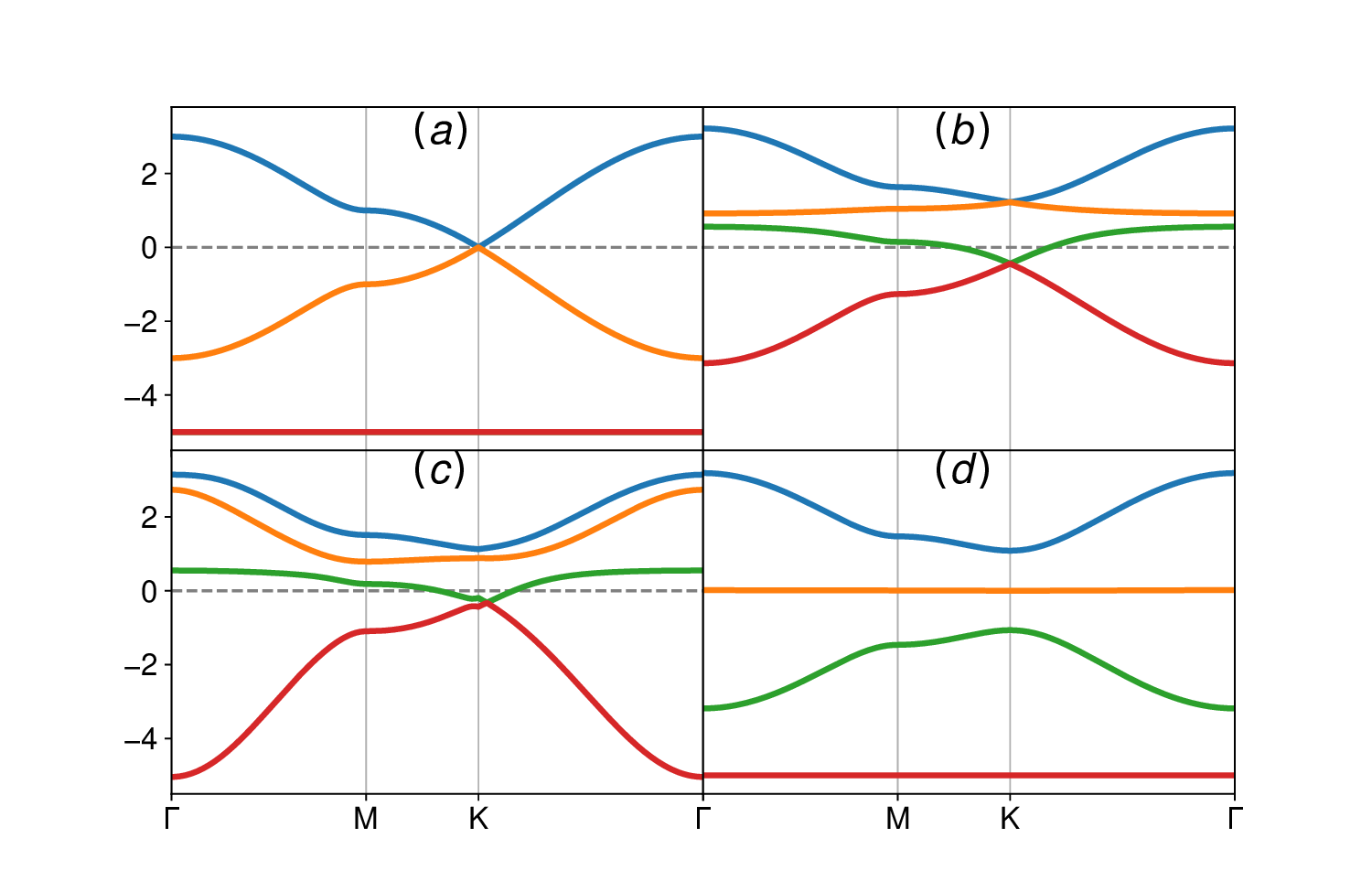}
\centering\includegraphics [scale=0.32]{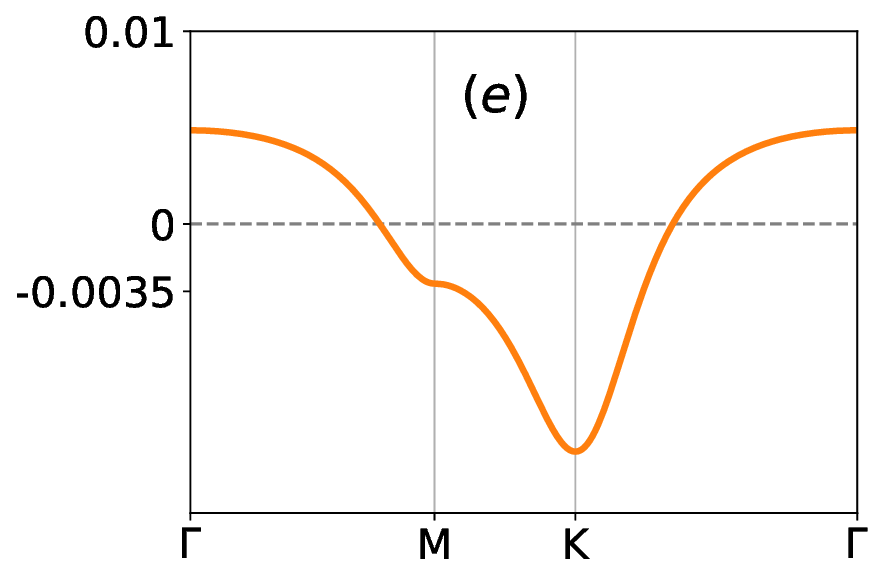}
\hspace*{-0.2cm}\includegraphics [scale=0.3]{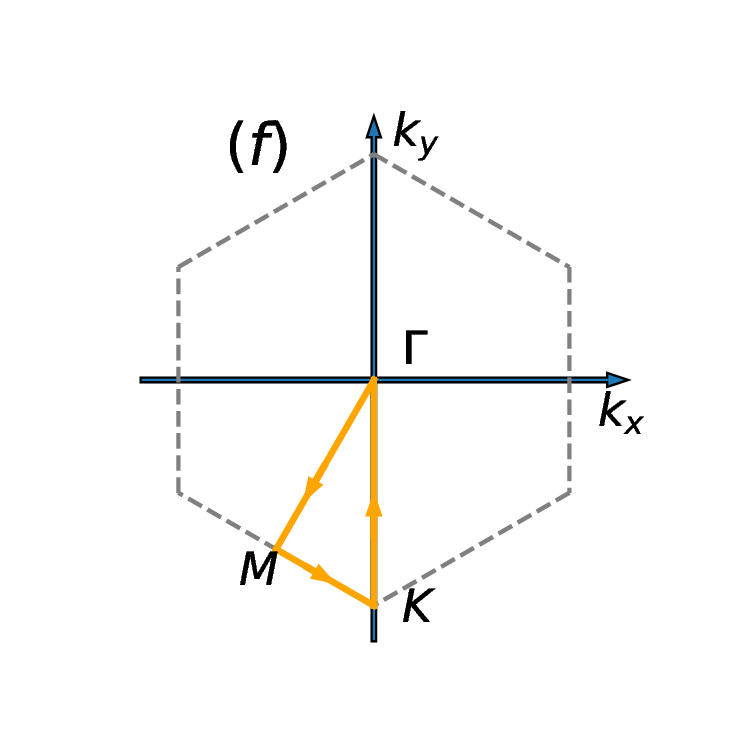}
\caption{\label{landscape}
Evolution of the band structure landscape: (a) The full-decoupled phase; (b) and (c) The full-KS phase with two identical or distinct Kondo scales with or without the inversion symmetry, respectively; (d) The genuine SKS with a single non-vanishing Kondo scale. Notice that the in-gap band is nearly flat. (e) The amplified band structure of the nearly flat band in the A-B pattern, with the band top at the $\Gamma$-point and bottom at the $K$-point. The Van Hove singularity is located at the $M$-point; (f) The plotted momentum path in the first Brillouin zone.
In the plots (a)-(e), the model parameters are $V=3,E_0=-5,t_f=0$ with $\beta=400$, the corresponding mean-field parameters as well as the chemical potential are all self-consistently solved based on the coupled mean-field equations. }
\end{figure}

{\it Landscape of band structures.}
 By tuning the sliding distance, a periodic evolution of such physics can be envisioned, and the successive transformations of the electronic band structures can be expected. As a function of the sliding distance and interlayer hybridization strength, the band structure exhibits a remarkable evolution as schematically illustrated in Fig. \ref{landscape}. There are four types of representative band structures in this correlated system  which are numerically solved based on the self-consistent mean-field equations.
 Here, Fig. \ref{landscape}(a) represents the band structure in the fully decoupled phase for the relatively small hybridization strength $V$; Fig. \ref{landscape}(b)/(c) represent the band structures in the full KS phase for sufficiently large $V$ with two identical or distinct Kondo scales with or without the inversion symmetry, respectively; Fig. \ref{landscape}(d) represents the band structure in the genuine SKS phase with a single non-vanishing Kondo scale. Notice that in this phase the second in-gap band closed to the Fermi energy is nearly flat. Notice also that the band at the bottom (the red line) represents the bare $f$ level (assumedly in the absence of the $f$ electron hopping and long-ranged interlayer hybridization, i.e., $t_f$=0, $\alpha=0$) as the bottom red line in Fig. \ref{landscape}(a), but complications may arise (e.g., it may shift close to the Fermi level and disperse in the SKS phase) in the presence of non-vanishing $t_f$, $\alpha$, and quantum fluctuations.

 Starting from the A-A pattern with a moderately large hybridization strength $V$,  we can envision that upon sliding, both the lower and upper Dirac bands (below or above the Kondo gap scaled by $r^2$) open a band gap (scaled by $\delta^2 x_0$) at the Dirac points (located at corners of the hexagonal BZ ) due to the inversion symmetry breaking similar to the situation in the multilayered graphene\cite{ManesPRB}. Near the A-B pattern, one of the bands could be decoupled in a wider parameter region. The bandwidth of the second band in the three hybridization bands is proportional to $r^2_{B}$ near the boundary of this phase, thus becoming nearly flat. The stability of the nearly flat band in this SKS phase is enhanced by the restored ${\cal C}_3$ symmetry.

{\it Phase transitions by tuning electric voltage.}
Given that the critical values of $V_{c,B}$ and $V_{c,A}$ are relatively large in our previous calculations, it is also necessary to briefly discuss the tunability of the SKS.
In the limit $U\rightarrow \infty$, the $f$ electrons are local magnetic moments as we assumed.  In the SKS phase, the effective interlayer Kondo coupling can be estimated by $J_{K} \approx \frac{2V^2}{|E_F-E_0|}$ in this limit, with $E_F$ is roughly around the Fermi energy and $E_0$ the energy level of $f$ electron. For the purpose of illustration, we have fixed $E_0=-5$ in most of our previous calculations and the resultant $V_{c,B}$ is about $1.5\sim 3.0$.  However, the relative energy difference $E_F-E_0$ can be tuned by applying the interlayer voltage $V_g$. For the homogeneous bilayer structure, this can effectively tune the electron densities in each layers. For the present heterostructure, it effectively shifts the interlayer energy difference to $E_F-E_0-V_g$, or equivalently $E_0\rightarrow \tilde E_0=E_0+V_g$. So increasing the voltage will increase $E_0$ and in turn increase the effective Kondo coupling $J_K$. As a comparison, we present the similar calculations for both $E_0=-5$ and $E_0=-3$ in Appendix C. It is shown that the critical $V_{c,B}$ is steadily reduced by decreasing the $f$ electron level $|E_0|$. Therefore, the SKS phase is accessible by applying the interlayer voltage and we expect that the successive KS transitions can be tuned accordingly.

{\it Implications for the general flat-band physics and strange metallicity.}
Finally, we note that the selective single-ion Kondo effect or other flat-band Kondo physics were discussed in the Lieb or kagome lattices \cite{tran2018molecular,2019Selective,kourris2023kondo}
where the conduction electron bath has both flat and dispersive bands by itself. By contrast, the flat-band and the associated strange metallicity discovered here are all the interaction-driven phenomena arising from the interplay between the lattice coherent Kondo effect and the inversion symmetry breaking. The strange metallicity associated with flat bands and Van Hove singularities has been an important theme in correlated electron systems, particularly in the graphene-based moire structures and transition-metal oxide layer structures. Recently, a unified framework for understanding the strange metallicity in these systems has been proposed based on the general Kondo or Anderson lattices, similar to the heavy fermion systems\cite{NRM2024, RamireaPRL,SongPRL,KumarPRB,Guerci, XiePRR}. In this framework, the $f$-orbitals come from the same type of conduction electrons but associated with much localized Wannier orbitals, or the so-called compact molecular orbitals representing some linear superpositions of atomic orbitals for different lattice sites \cite{ChenNC, HuSA,Mahankali}. The quantum criticality associated with the Kondo destruction provides a reasonable mechanism for the strange metal behaviors such as linear-in-temperature electric resistivity, dynamical Plankian scaling, and jump of Fermi surface in these systems. Again, it is interesting to notice that the flat band discovered in the present work is due to the hybridization between the different types of electrons, i.e., itinerant orbitals and purely-localized $f$ orbitals. It exists in the SKS phase even at half-filling and develops only in the Kondo gap. Specifically, it disappears upon Kondo destruction, resulting in a jump in the Fermi volume or the reconstruction of the Fermi surface crossing the critical point $V_{c,B}$, violating Luttinger theorem. The appearance of a Van Hove singularity at the Fermi energy in this flat-band SKS phase could also result in ferromagnetism and other correlation effects.
 Moreover, inclusion of the spin-orbit interactions in both the host electrons and the interlayer hybridizations could lead to the topological Kondo insulator and other topological semimetallic phases in the related Anderson lattice systems\cite{FengPRL, lu2019tunable}.

Therefore, our results add a new ingredient to a wide context of the flat band physics and Kondo physics in correlated electron systems. We expect that the correlated bilayers like the transition metal-dichalcogenide heterostructures and the densely $f$ electron doped graphene bilayers are suitable platforms to observe these fascinating quantum phenomena.

\acknowledgments

The authors thank Y. Liu and H.Q. Yuan for useful discussions. This work was supported in part by the National Science
Foundation of China under Grants No. 12274109 and 12274364.

\begin{appendix}

\section{Interlayer hybridization cut-off schemes}

\subsection{Distance between different sites }

The heterostructure  we are studying consists of a $c$-layer and a $f$-layer, with the same honeycomb lattice.
The $c$-layer is put on top of the $f$-layer. The position vector of any site on the $f$-layer is denoted as
\begin{equation}
\vec R_{{\bf n},\eta}=n_{1}\vec a_{1}+n_{2}\vec a_{2}+\frac{a_0}{2}\epsilon_{\eta}\vec e_{x}
\end{equation}
with $\epsilon_{\eta}=\pm1$ for $\eta=B/A$ sublattices, $a_0$ being the distance between the nearest-neighbor sites,
and $\vec a_{1/2}=\frac{3a_0}{2}\vec e_{x}\pm\frac{\sqrt{3}a_0}{2}\vec e_{y}$.
The c-layer is shifted by $\delta x_0=a_0 \delta x$ along the armchair direction (chosen as the $x$-axis), with $\delta x$ being a dimensionless parameter measuring the relative shift between the two layers. The position vector of any site
on the c-layer is represented by
\begin{equation}
\vec r_{{\bf m},\eta}=m_{1}\vec a_{1}+m_{2} \vec a_{2}+\frac{a_0}{2}\epsilon_{\eta} \vec e_{x}+\delta x_0 \vec e_{x}+a_{z}\vec e_{z}~.
\end{equation}
In the following, all site-site distances are measured with respect to $a_0$ which is served as the length unit (so that $\delta x_0=\delta x$ throughout the Appendices).

\begin{figure}
  \includegraphics [scale=1.4]{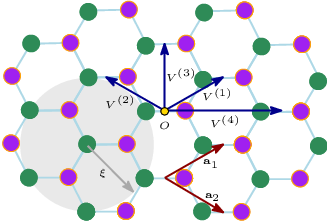}
  \caption{\label{fig:Hopping-matrix-elements}Hybridization matrix elements between
the origin of the $c$-layer (denoted as $O$) and different neighboring
cell of the $f$-layer, represented as the blue lines with arrows. The
origin for the $c$-layer will move along the $x$-axis while the $f$-layer
and thus target positions (location of arrow) are unchanged, then
the distances will change accordingly. The hybridization matrix elements
are cut-off by planar distance $\xi$ (say $\sqrt{3}a$) shown in
as the gray circle in the figure.}
\end{figure}

The distance of a site on the $f$-layer to another site on the $c$-layer is
\begin{align}
  & d_{ \vec r_{{\bf m},\eta}(\delta x)\leftarrow
  \vec R_{{\bf n},\eta^{\prime}}}\nonumber\\
\equiv & | \vec r_{{\bf m},\eta}- \vec R_{{\bf n},\eta^{\prime}}|\nonumber\\
= & |(m_{1}-n_{1}) \vec a_{1}+(m_{2}-n_{2})\vec a_{2}+[\frac{1}{2}(\epsilon_{\eta}-
\epsilon_{\eta^{\prime}}
)+\delta x] \vec e_{x}+a_{z} \vec e_{z}|\nonumber\\
	= & \{ [\frac{3}{2}(m_{1}-n_{1})+\frac{3}{2}(m_{2}-n_{2})+\frac{1}{2}(\epsilon_{\eta}-
\epsilon_{\eta^{\prime}}
)\nonumber\\
	&+\delta x]^{2}+[\frac{\sqrt{3}}{2}(m_{1}-n_{1})-\frac{\sqrt{3}}{2}(m_{2}-n_{2})]^{2}+a_{z}^{2}\}^{1/2}
\end{align}
Since the interlayer hybridization matrix elements depend on the relative distance between
the sites on both $f$-layer and $c$-layer, we might fix the unit-cell of $c$-layer at the origin (namely ${\bf m}=0$) to simplify the expressions. Therefore, we have (see
Fig. \ref{fig:Hopping-matrix-elements} for the most
relevant hopping terms)
\begin{align}
       & d_{{\bf n};\delta x;\eta,\eta^{\prime}}\nonumber\\
\equiv & d_{ \vec r_{{\bf m}=0,\eta}(\delta x)\leftarrow \vec R_{{\bf n},\eta^{\prime}}}\nonumber\\
     = & \frac{1}{2}\sqrt{[-3(n_{1}+n_{2})+(\epsilon_{\eta}-
\epsilon_{\eta^{\prime}}
)+2\delta x]^{2}+3(n_{1}-n_{2})^{2}+4a_{z}^{2}}
\end{align}

Considering the interlayer hybridization elements decay quickly as the distance increases,
we could focus on the major elements which connecting the nearest neighboring sites and so on within each layers. The first one
is the intra-cell hybridization element which is a function of distance
between the two-sites on the two layers
\begin{equation}
V_{{\bf n}=0;\delta x;\eta,\eta^{\prime}}=F(d{}_{{\bf n}=0;\delta x;\eta,\eta^{\prime}})\equiv V_{\eta,\eta^{\prime}}^{(0)}
\end{equation}
with
\begin{equation}
d_{{\bf n}=(0,0);\delta x;\eta,\eta^{\prime}}=\frac{1}{2}\sqrt{[
(\epsilon_{\eta}-
\epsilon_{\eta^{\prime}}
)+2\delta x]^{2}+4a_{z}^{2}}\equiv d^{(0)}
\end{equation}
The hybridization elements from the origin to its nearest neighbors have
following terms
\begin{equation}
V_{(1,0);\delta x;\eta,\eta^{\prime}}=V_{(0,1);\delta x;\eta,\eta^{\prime}}=F(d{}_{{\bf n}=(1,0);\delta x;\eta,\eta^{\prime}})\equiv V_{\eta,\eta^{\prime}}^{(1)}
\end{equation}
\begin{equation}
V_{(-1,0);\delta x;\eta,\eta^{\prime}}=V_{(0,-1);\delta x;\eta,\eta^{\prime}}=F(d{}_{{\bf n}=(-1,0);\delta x;\eta,\eta^{\prime}})\equiv V_{\eta,\eta^{\prime}}^{(2)}
\end{equation}
\begin{equation}
V_{(1,-1);\delta x;\eta,\eta^{\prime}}=V_{(-1,1);\delta x;\eta,\eta^{\prime}}=F(d{}_{{\bf n}=(1,-1);\delta x;\eta,\eta^{\prime}})\equiv V_{\eta,\eta^{\prime}}^{(3)}
\end{equation}
since
\begin{align}
	& d_{{\bf n}=(1,0);\delta x;\eta,\eta^{\prime}}=d_{{\bf n}=(0,1);\delta x;\eta,\eta^{\prime}}\nonumber\\
	=& \frac{1}{2}\sqrt{[-3+(\epsilon_{\eta}-
\epsilon_{\eta^{\prime}}
)+2\delta x]^{2}+3+4a_{z}^{2}}\equiv d^{(1)}
\end{align}

\begin{align}
	&d_{{\bf n}=(-1,0);\delta x;\eta,\eta^{\prime}}=d_{{\bf n}=(0,-1);\eta,\eta^{\prime}}\nonumber\\
	=&\frac{1}{2}\sqrt{[3+(\epsilon_{\eta}-
\epsilon_{\eta^{\prime}}
)+2\delta x]^{2}+3+4a_{z}^{2}}\equiv d^{(2)}
\end{align}

\begin{align}
	&d_{{\bf n}=(1,-1);\delta x;\eta,\eta^{\prime}}=d_{{\bf n}=(-1,1);\delta x;\eta,\eta^{\prime}}\nonumber\\
	=&\frac{1}{2}\sqrt{[
 (\epsilon_{\eta}-
\epsilon_{\eta^{\prime}}
)+2\delta x]^{2}+12+4a_{z}^{2}}\equiv d^{(3)}
\end{align}
In addition, we need to consider one hybridization element for the next nearest neighbor sites since
we are moving the $c$-layer along the $x$-axis
\begin{equation}
V_{(1,1);\delta x;\eta,\eta^{\prime}}=F(d{}_{{\bf n}=(1,1);\delta x;\eta,\eta^{\prime}})\equiv V_{\eta,\eta^{\prime}}^{(4)}
\end{equation}
with
\begin{equation}
d_{n=(1,1);\delta x;\eta,\eta^{\prime}}=\frac{1}{2}\sqrt{[-6+(\epsilon_{\eta}-
\epsilon_{\eta^{\prime}}
)+2\delta x]^{2}+4a_{z}^{2}}\equiv d^{(4)}
\end{equation}

\subsection{Approximation to the hybridization matrix elements}

\begin{figure}
\hspace*{-0.6cm}\centering\includegraphics[scale=0.5]{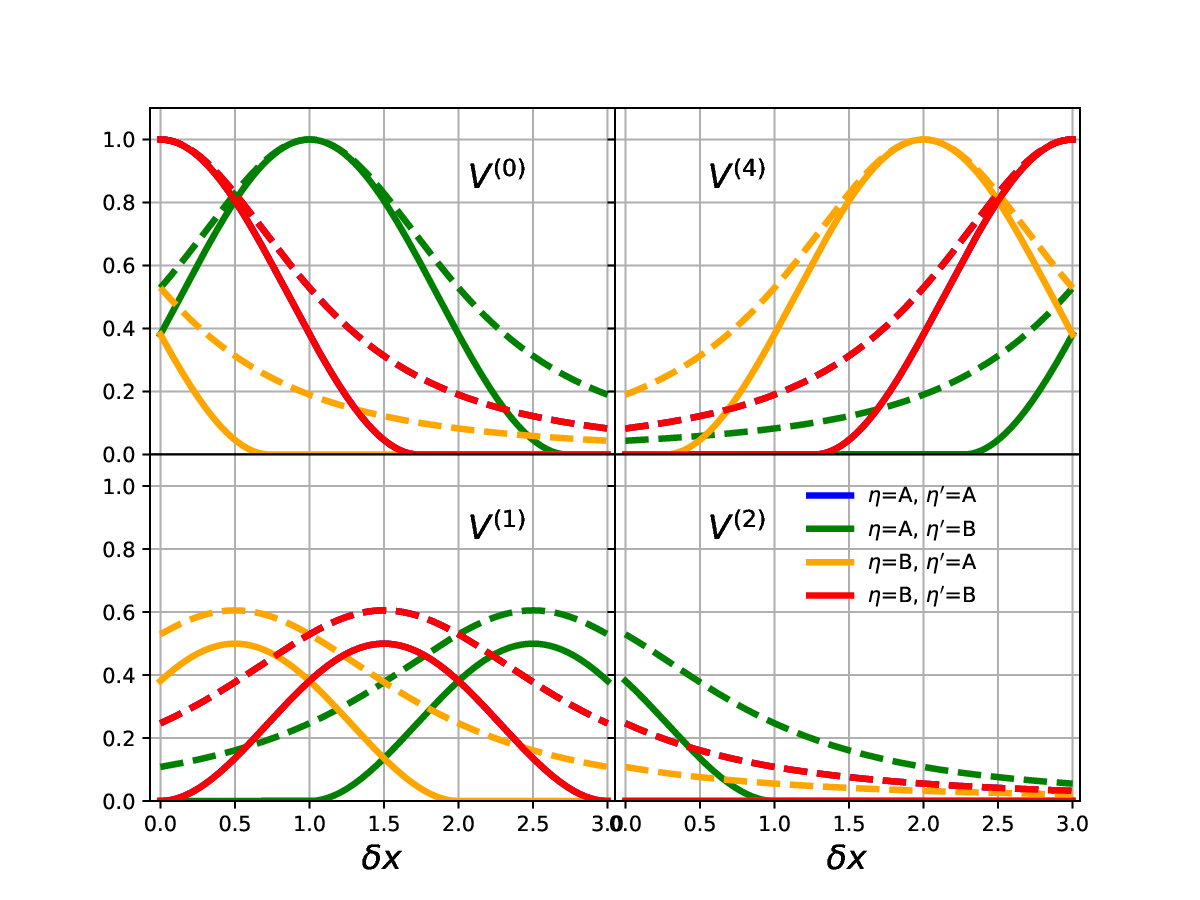}\caption{\label{fig:hopping-NNN-cutoff}$\delta x$-dependence of different hybridization elements with $a_z=1.5 a_0$, $\zeta=2$ using the the decay cut-off function (dashed lines) and its reparameterized version  (solid lines). Note that the cutoff radius is chosen
as $d_{\max}=\sqrt{a_{z}^{2}+3a_0^{2}}$ (such that those next nearest
neighboring (NNN) sites are on the cutoff sphere) is used. Note that only $V^{(3)}_{\eta,\eta'}$ (not shown) vanishes within
such cutoff radius. Also note that the two components $AA$ and $BB$
are the same.}
\end{figure}

\begin{figure}
\hspace*{-0.6cm}\centering\includegraphics[scale=0.5]{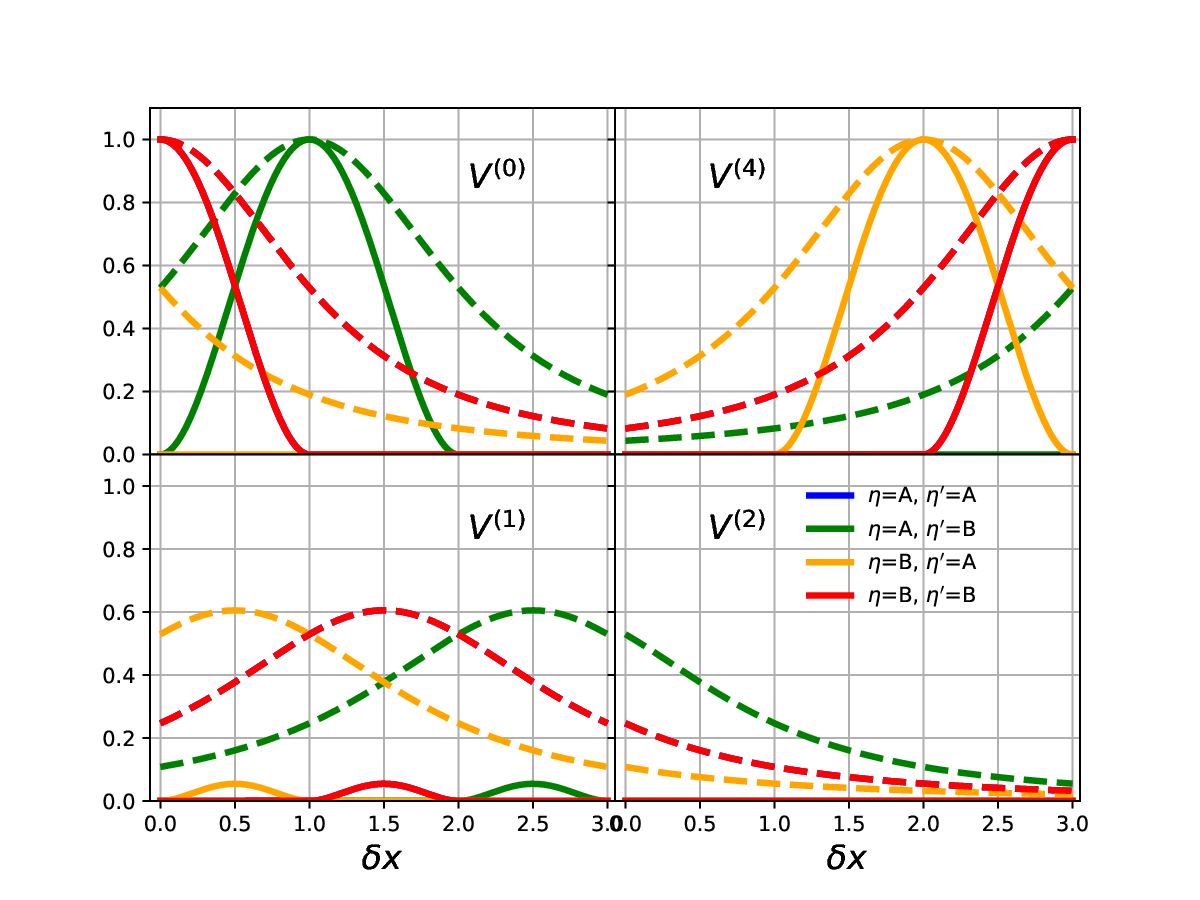}
\caption{\label{fig:hopping-NN-cutoff}$\delta x$-dependence of different
hybridization elements with $a_z=1.5 a_0$, $\zeta=2$ using the decay cut-off function (dashed lines) and its reparameterized version (solid lines). Note that the cutoff radius is chosen
as $d_{\max}=\sqrt{a_{z}^{2}+a_0^{2}}$ (such that those nearest neighboring
(NN) sites are on the cutoff sphere). Note that both $V^{(2)}_{\eta,\eta'}$ and $V^{(3)}_{\eta,\eta'}$ (not shown) vanish within
such cutoff radius. Also note that the two components $AA$ and $BB$
are the same.}
\end{figure}

 Each interlayer hybridization matrix elements between two atomic sites are defined as the overlaps of their local atomic orbital wavefunctions. Besides the orbital character, they depend mainly on the distance between the two sites and can be roughly parameterized by a decaying function
\begin{equation}
V(d)=V[\frac{a_{z}}{d}]^{\zeta}e^{-|d-a_{z}|/\xi}~.
\end{equation}
As explained in the main text, we could choose $V=V_{1}$ as a single tuning parameter for a given distance $d$, while $\zeta$ and $\xi$ are simply implemented by some cutoff schemes.
The cutoff scheme we adopted here assumes that the hybridization element decays smoothly and vanishes
beyond some chosen cutoff spheres. For this purpose, we use the following reparameterization version of the decaying function
\begin{equation}
V(d)=V\theta(d_{\max}-d)(\frac{d_{\max}-a_{z}}{d_{\max}-d})^{\zeta}~,
\end{equation}
where $\theta(x)$ is the step function, and $d_{max}\sim\xi$ the radius of
the cutoff sphere. It could be either
\begin{equation}
\xi\sim d_{\max}=\sqrt{a_{z}^{2}+a_0^{2}}
\end{equation}
where the cutoff radius is within the nearest neighboring (NN) sites
on each layer, or
\begin{equation}
\xi \sim d_{\max}=\sqrt{a_{z}^{2}+3a_0^{2}}
\end{equation}
where the cutoff radius is within the next nearest neighboring (NNN) sites. The reparameterized version of the decaying function is more convenient in numerical simulation.
A comparison between different cutoff schemes is shown in the figures (Fig. \ref{fig:hopping-NNN-cutoff}
and Fig. \ref{fig:hopping-NN-cutoff}) using $a_z=1.5 a_0$, $\zeta=2$. In these figures $V^{(i)}_{\eta,\eta'}$ are the intra-cell ($i=0$) and inter-cell ($i=1,2,3,4$) hybridization elements defined in Appendix A.1. Not confused with those denoted by $V_1$ or $V_2$ in the main text, these hybridization elements vary with the sliding distance without a clear distinction between the nearest neighbor (NN) or the next-nearest-neighbor (NNN) ones for arbitrary $\delta x_0$ except for $\delta x_0\sim 0, a_0$ where the distinction becomes robust. The dashed or solid lines in these plots correspond to the schemes with the the decaying cut-off function or its reparameterized version, respectively. Therefore we observe in both cases that it is sufficient to take into account the NN and NNN interlayer hybridization elements in the vicinity of $\delta x_0=0, a_0$, respectively.

\section{Eigenvalue problem of the Hamiltonian matrix }

\subsection{Hamiltonian matrix elements}
\begin{figure}
  \includegraphics [scale=0.4]{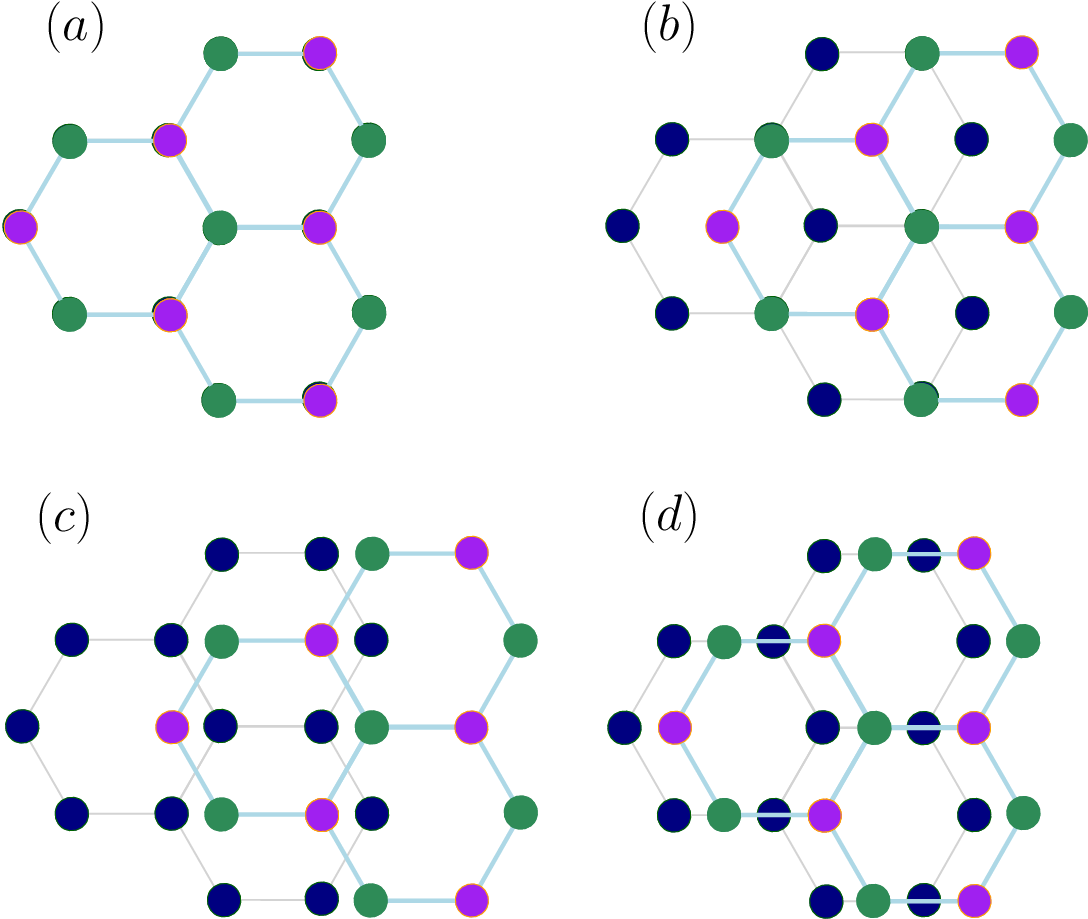}
  \caption{\label{fig:lattice_shift}
  Four shift patterns with relatively high symmetries: a), $\delta x=0$:
  A-A stack pattern, b), $\delta x=1$: A-B stack pattern,
  c), $\delta x=1.5$: M stack pattern, and  d), $\delta x=0.5$. }
\end{figure}

Using the Fourier transformation
\begin{equation}
\hat c_{\vec r\sigma} = \frac{1}{\sqrt{N}} \sum_{\bf{k}} e^{ i{\bf k}\cdot \vec{r}} \hat c_{\bf{k}\sigma},
\end{equation}
the Hamiltonian matrix elements can be expressed as
\begin{align}
	&	h_{\eta,\eta^{\prime}}({\bf k})\nonumber\\
	= & e^{-i k_x [\frac{1}{2}(\epsilon_{\eta}-
\epsilon_{\eta^{\prime}}
)+\delta x]}\nonumber\\
	  & \times[V_{\eta,\eta^{\prime}}^{(0)}+(e^{i{\bf k}\cdot \vec a_{1}}+e^{i{\bf k}\cdot \vec a_{2}})V_{\eta,\eta^{\prime}}^{(1)}+(e^{-i{\bf k}\cdot \vec a_{1}}+e^{-i{\bf k}\cdot \vec a_{2}})V_{\eta,\eta^{\prime}}^{(2)}\nonumber\\
	& +(e^{i{\bf k}\cdot( \vec a_{1}- \vec a_{2})}+e^{-i{\bf k}\cdot(\vec a_{1}-\vec a_{2})})V_{\eta,\eta^{\prime}}^{(3)}+e^{i{\bf k}\cdot( \vec a_{1}+ \vec a_{2})}V_{\eta,\eta^{\prime}}^{(4)}]
\end{align}
with
\begin{equation}
f_{{\bf k}}=e^{ik_x}(1+e^{-i{\bf k}\cdot \vec a_{1}}+e^{-i{\bf k}\cdot \vec a_{2}})~.
\end{equation}
Then, using the slave-boson method, the Hamiltonian matrix in $\bf k$-space can be written as
\begin{equation}
H_{{\bf k}\sigma}=\begin{bmatrix}0 & -tf_{{\bf k}} & r_{A}h_{AA} & r_{B}h_{AB}\\
-tf_{{\bf k}}^{*} & 0 & r_{A}h_{BA} & r_{B}h_{BB}\\
r_{A}h_{AA}^{*} & r_{A}h_{BA}^{*} & E_{0}+\lambda_{A} & -t_{f}r_{A}r_{B}f_{{\bf k}}\\
r_{B}h_{AB}^{*} & r_{B}h_{BB}^{*} & -t_{f}r_{A}r_{B}f_{{\bf k}}^{*} & E_{0}+\lambda_{B}
\end{bmatrix}
\end{equation}
In order to numerically solve problem efficiently, the above Hamiltonian is rewritten as two-by-two block
matrix
\begin{equation}
H_{{\bf k}\sigma}=\begin{bmatrix}H_{cc} & H_{cf}\\
H_{fc} & H_{ff}
\end{bmatrix}
\end{equation}
The two diagonal block elements are
\begin{equation}
H_{cc}=\begin{bmatrix}0 & -tf_{{\bf k}}\\
-tf_{{\bf k}}^{*} & 0
\end{bmatrix}
\end{equation}
and
\begin{equation}
H_{ff}=\begin{bmatrix}E_{0}+\lambda_{A} & -t_{f}r_{A}r_{B}f_{{\bf k}}\\
-t_{f}r_{A}r_{B}f_{{\bf k}}^{*} & E_{0}+\lambda_{B}
\end{bmatrix}
\end{equation}
for the $c$-layer and $f$-layer, respectively. They remain unchanged during the
shifting. While the off-diagonal blocks $H_{fc}=H_{cf}^{*}$ take the form
\begin{equation}
H_{cf}=\begin{bmatrix}r_{A}h_{AA} & r_{B}h_{AB}\\
r_{A}h_{BA} & r_{B}h_{BB}
\end{bmatrix}~.
\end{equation}
For several special shift points (see Fig. \ref{fig:lattice_shift}),
we have simpler expressions for this sub-matrix as detailed in the following.

\subsubsection{$\delta x=0$}
This is the A-A stack pattern:
\begin{equation}
H_{cf}^{\delta x=0}=\begin{bmatrix}r_{A}V_{1} & r_{B}V_{2}f_{{\bf k}}\\
r_{A}V_{2}f_{{\bf k}}^{*} & r_{B}V_{1}
\end{bmatrix}.
\end{equation}

\subsubsection{$\delta x=1$}
This is the A-B stack pattern:

\begin{equation}
H_{cf}^{\delta x=1}=\begin{bmatrix}r_{A}V_{2}f_{{\bf k}}^{*} & r_{B}V_{1}\\
r_{A}V_{2}f_{{\bf k}} & r_{B}V_{2}f_{{\bf k}}^{*}
\end{bmatrix}.
\end{equation}

\subsubsection{$\delta x=1.5$}
This is the M stack pattern:

\begin{equation}
H_{cf}^{\delta x=1.5}=\begin{bmatrix}r_{A}V_{2}g_{1} & r_{B}V_{1}g_{2}\\
r_{A}V_{1}g_{2}^{*} & r_{B}V_{2}g_{1}
\end{bmatrix},
\end{equation}
with
\begin{equation}
g_{1}=e^{-i\frac{3}{2}k_x}[e^{i{\bf k}\cdot \vec a_{1}}+e^{i{\bf k}\cdot \vec a_{2}}]
\end{equation}
and
\begin{equation}
g_{2}=e^{-i\frac{1}{2}k_x}~.
\end{equation}

\subsubsection{$\delta x=0.5$}
This pattern is not specified in the main text. In this case, we have
\begin{equation}
H_{cf}^{\delta x=0.5}=\begin{bmatrix}r_{A}V_{1}g_2 & r_{B}V_{1}g_2^*\\
r_{A}V_{2}g_1 & r_{B}V_{1}g_2
\end{bmatrix},
\end{equation}
with $g_1$ and $g_2$ being defined as in the above.

\subsection{Solve the four-band model}

Let $H_{4\times 4, {\bf k\sigma}}$ be $4\times 4$ Hamiltonian matrix with eigenvalue $y$.
By requiring that
\begin{equation}
\det(y I_{4\times 4}-H_{4\times 4, {\bf k}\sigma})=0~,
\end{equation}
we have
\begin{equation}
y^{4}+by^{3}+cy^{2}+dy+e=0~,
\end{equation}
with the coefficients being given by
\begin{equation}
b=-(E_{0}+\lambda_{A})-(E_{0}+\lambda_{B}),
\end{equation}
\begin{align}
c & =(E_{0}+\lambda_{A})(E_{0}+\lambda_{B})-(1+t_{f}^{2}r_{A}^{2}r_{B}^{2})\epsilon_{k}^{2}-r_{A}^{2}g_{1k}-r_{B}^{2}g_{2k}~,
\end{align}
\begin{align}
d= &r_{A}^{2}[(E_{0}+\lambda_{B})g_{1k}+g_{3k}]+r_{B}^{2}[(E_{0}+\lambda_{A})g_{2k}+g_{4k}]\nonumber\\
+ & t_{f}r_{A}^{2}r_{B}^{2}g_{5k}+[(E_{0}+\lambda_{A})+(E_{0}+\lambda_{B})]\epsilon_{k}^{2}~,
\end{align}
\begin{align}
e= &  r_{A}^{2}r_{B}^{2}g_{6k}-t_{f}r_{A}^{2}r_{B}^{2}g_{7k}-(E_{0}+\lambda_{B})r_{A}^{2}g_{3k}\nonumber\\
-&(E_{0}+\lambda_{A})r_{B}^{2}g_{4k}
-  (E_{0}+\lambda_{A})(E_{0}+\lambda_{B})\epsilon_{k}^{2}+t_{f}^{2}r_{A}^{2}r_{B}^{2}\epsilon_{k}^{4}~,
\end{align}
in which $\epsilon_{k}\equiv|t_c f_{\bf k}|$, and
\begin{equation}
g_{1k}\equiv|h_{AA}|^{2}+|h_{BA}|^{2}~,
\end{equation}
\begin{equation}
g_{2k}\equiv|h_{AB}|^{2}+|h_{BB}|^{2}~,
\end{equation}
\begin{equation}
g_{3k}\equiv fh_{AA}^{*}h_{BA}+f^{*}h_{AA}h_{BA}^{*}~,
\end{equation}
\begin{equation}
g_{4k}\equiv fh_{AB}^{*}h_{BB}+f^{*}h_{AB}h_{BB}^{*}~,
\end{equation}
\begin{align}
g_{5k}\equiv& fh_{AA}h_{AB}^{*}+fh_{BA}h_{BB}^{*}\nonumber\\
+&f^{*}h_{AA}^{*}h_{AB}+f^{*}h_{BA}^{*}h_{BB}~,
\end{align}
\begin{align}
g_{6k}\equiv&|h_{AB}|^{2}|h_{BA}|^{2}+|h_{AA}|^{2}|h_{BB}|^{2}\nonumber\\
-&h_{AA}h_{BB}h_{AB}^{*}h_{BA}^{*}-h_{AA}^{*}h_{BB}^{*}h_{AB}h_{BA}~,
\end{align}
\begin{align}
g_{7k}\equiv&(h_{AB}^{*}h_{BA}f^{2}+h_{AB}h_{BA}^{*}f^{*2})\nonumber\\
+&(h_{AA}^{*}h_{BB}+h_{AA}h_{BB}^{*})\epsilon_{k}^{2}~.
\end{align}

The four roots $y_m$ ( the band energies which were denoted by $E_{\textbf k \sigma m}$ in the main text for $m=1, 2, 3, 4$) can be formally solved as \cite{cardano2007rules}:
\begin{align}
y_m=\frac{-b+(-1)^{m/2}M+(-1)^{m+1}\sqrt{S+(-1)^{m/2}T}}{4},
\end{align}
with
\begin{align*}
&M=\sqrt{b^2-\frac{8}{3}c+4\left( \omega^{k-1}u+\omega^{4-k}v\right)},\\
&S=2b^2-\frac{16}{3}c-4\left( \omega^{k-1}u+\omega^{4-k}v\right),\\
&T=\frac{8bc-16d-2b^3}{M}.
\end{align*}
In above, $k$ takes integers $\{1, 2, 3\}$ such that the absolute value of $M$ is maximized among the possible choices for $k$;
$\omega=e^{i \frac{2\pi}{3}}$ and
\begin{align*}
u &= \sqrt{Q + D}, \\
v &= \sqrt{Q - D}, \\
Q &= \frac{27d^2 + 2c^3 + 27b^2e - 72ce - 9bcd}{54}, \\
D &= \sqrt{Q^2 - P^3}, \\
P &= \frac{c^2 + 12e - 3bd}{9}. \\
\end{align*}

Although we can use the formal algebraic solutions of the $4\times 4$ Hamiltonian matrix, this approach is not efficient in our mean-field-based numerical calculations. In the mean-field calculations, we need different kinds of derivations of the band energy (the $\bf k$-dependent eigenvalue $y$) with respect to the mean-field parameters. Most of them depend on
the shift distance $\delta x_0$. These derivatives are listed below:
\begin{equation}
\frac{\partial y}{\partial\lambda_{\alpha}}=\frac{1}{\Delta}(\frac{\partial b}{\partial\lambda_{\alpha}}y^{3}+\frac{\partial c}{\partial\lambda_{\alpha}}y^{2}+\frac{\partial d}{\partial\lambda_{\alpha}}y+\frac{\partial e}{\partial\lambda_{\alpha}})~,
\end{equation}
\begin{equation}
\frac{1}{2r}\frac{\partial y}{\partial r_{\alpha}}=\frac{1}{\Delta}(\frac{1}{2r}\frac{\partial c}{\partial r_{\alpha}}y^{2}+\frac{1}{2r}\frac{\partial d}{\partial r_{\alpha}}y+\frac{1}{2r}\frac{\partial e}{\partial r_{\alpha}})~,
\end{equation}
with
\begin{equation}
\Delta=-(4y^{3}+3by^{2}+2cy+d)~,
\end{equation}
\begin{equation}
\frac{\partial b}{\partial\lambda_{\alpha}}=-1~,
\end{equation}
\begin{equation}
\frac{1}{2r}\frac{\partial b}{\partial r_{\alpha}}=0~,
\end{equation}
\begin{equation}
\frac{\partial c}{\partial\lambda_{A}}=(E_{0}+\lambda_{B})~,
\end{equation}
\begin{equation}
\frac{\partial c}{\partial\lambda_{B}}=(E_{0}+\lambda_{A})~,
\end{equation}
\begin{equation}
\frac{1}{2r_{A}}\frac{\partial c}{\partial r_{A}}=-t_{f}^{2}r_{B}^{2}\epsilon_{k}^{2}-g_{1k}~,
\end{equation}
\begin{equation}
\frac{1}{2r_{B}}\frac{\partial c}{\partial r_{B}}=-t_{f}^{2}r_{A}^{2}\epsilon_{k}^{2}-g_{2k}~,
\end{equation}
\begin{equation}
\frac{\partial d}{\partial\lambda_{A}}=r_{B}^{2}g_{2k}+\epsilon_{k}^{2}~,
\end{equation}
\begin{equation}
\frac{\partial d}{\partial\lambda_{B}}=r_{A}^{2}g_{1k}+\epsilon_{k}^{2}~,
\end{equation}
\begin{equation}
\frac{1}{2r_{A}}\frac{\partial d}{\partial r_{A}}=(E_{0}+\lambda_{B})g_{1k}+g_{3k}+t_{f}r_{B}^{2}g_{5k}~,
\end{equation}
\begin{equation}
\frac{1}{2r_{B}}\frac{\partial d}{\partial r_{B}}=(E_{0}+\lambda_{A})g_{2k}+g_{4k}+t_{f}r_{A}^{2}g_{5k}~,
\end{equation}
\begin{equation}
\frac{\partial e}{\partial\lambda_{A}}=-r_{B}^{2}g_{4k}-(E_{0}+\lambda_{B})\epsilon_{k}^{2}~,
\end{equation}
\begin{equation}
\frac{\partial e}{\partial\lambda_{B}}=-r_{A}^{2}g_{3k}-(E_{0}+\lambda_{A})\epsilon_{k}^{2}~,
\end{equation}
\begin{equation}
\frac{1}{2r_{A}}\frac{\partial e}{\partial r_{A}}=r_{B}^{2}g_{6k}-t_{f}r_{B}^{2}g_{7k}-(E_{0}+\lambda_{B})g_{3k}+t_{f}^{2}r_{B}^{2}\epsilon_{k}^{4}~,
\end{equation}
\begin{equation}
\frac{1}{2r_{B}}\frac{\partial e}{\partial r_{B}}=r_{A}^{2}g_{6k}-t_{f}r_{A}^{2}g_{7k}-(E_{0}+\lambda_{A})g_{4k}+t_{f}^{2}r_{A}^{2}\epsilon_{k}^{4}~.
\end{equation}

\subsection{Solve the three-band model}

In the SKS phase, $r_{A}$ is zero, the $f_{A}$-sublattice is essentially decoupled
from the rest of the system and thus we might use the three-band model $H_{3\times 3, {\bf k}\sigma}$ in calculations. By requiring that
\begin{equation}
\det(y I_{3\times 3}-H_{3\times 3, {\bf k}\sigma})=0~,
\end{equation}
the eigenvalue $y$ satisfies
\begin{equation}
y^{3}+by^{2}+cy+d=0~,
\end{equation}
with
\begin{equation}
b=-(E_{0}+\lambda_{B})~,
\end{equation}
\begin{equation}
c=-\epsilon_{k}^{2}-r_{B}^{2}g_{2k}~,
\end{equation}
\begin{equation}
d=(E_{0}+\lambda_{B})\epsilon_{k}^{2}+r_{B}^{2}g_{4k}~,
\end{equation}
and
\begin{equation}
g_{2k}\equiv|h_{AB}|^{2}+|h_{BB}|^{2}~,
\end{equation}
\begin{equation}
g_{4k}\equiv fh_{AB}^{*}h_{BB}+f^{*}h_{AB}h_{BB}^{*}~,
\end{equation}

As derived from \cite{cardano2007rules,lambert1906generalized}, the three roots can be formally solved as
\begin{equation}
\begin{aligned}
y_1 &= -\frac{b}{3}+2\sqrt[3]{r} \cos \theta, \\
y_2 &= -\frac{b}{3}+2\sqrt[3]{r} \cos(\theta + \frac{2\pi}{3}), \\
y_3 &=-\frac{b}{3} +2\sqrt[3]{r} \cos(\theta - \frac{2\pi}{3} ),
\end{aligned}
\end{equation}
where
\begin{equation}
\begin{aligned}
r& = \sqrt{-\left(\frac{p}{3}\right)^3}, \quad \theta = \frac{1}{3}\arccos\left(-\frac{q}{2r}\right), \quad p = \frac{3c-b^2}{3},\\
q& = \frac{27d-9bc+2b^3}{27}.\nonumber
\end{aligned}
\end{equation}

Similar to the four-band model, we need the following derivatives to solve the corresponding mean-field equations (the subscript $B$ for $f_B$-sublattice is implied for simplicity):
\begin{equation}
\frac{\partial y}{\partial\lambda}=\frac{1}{\Delta}(\frac{\partial b}{\partial\lambda}y^{2}+\frac{\partial c}{\partial\lambda}y+\frac{\partial d}{\partial\lambda})~,
\end{equation}
\begin{equation}
\frac{1}{2r}\frac{\partial y}{\partial r}=\frac{1}{\Delta}(\frac{1}{2r}\frac{\partial c}{\partial r}y+\frac{1}{2r}\frac{\partial d}{\partial r})~,
\end{equation}
with
\begin{equation}
\Delta=-(3y^{2}+2by+c)~,
\end{equation}
\begin{equation}
\frac{\partial b}{\partial\lambda}=-1~,
\end{equation}
\begin{equation}
\frac{1}{2r}\frac{\partial b}{\partial r}=0~,
\end{equation}
\begin{equation}
\frac{\partial c}{\partial\lambda}=0~,
\end{equation}
\begin{equation}
\frac{1}{2r}\frac{\partial c}{\partial r}=-(|h_{AB}|^{2}+|h_{BB}|^{2})~,
\end{equation}
\begin{equation}
\frac{\partial d}{\partial\lambda}=\epsilon_{k}^{2}~,
\end{equation}
\begin{equation}
\frac{1}{2r}\frac{\partial d}{\partial r}=g_{4k}~.
\end{equation}

\section{The mean-field solutions using different cutoff schemes and $f$-electron levels }

\begin{figure}
\centering\includegraphics[scale=0.45]{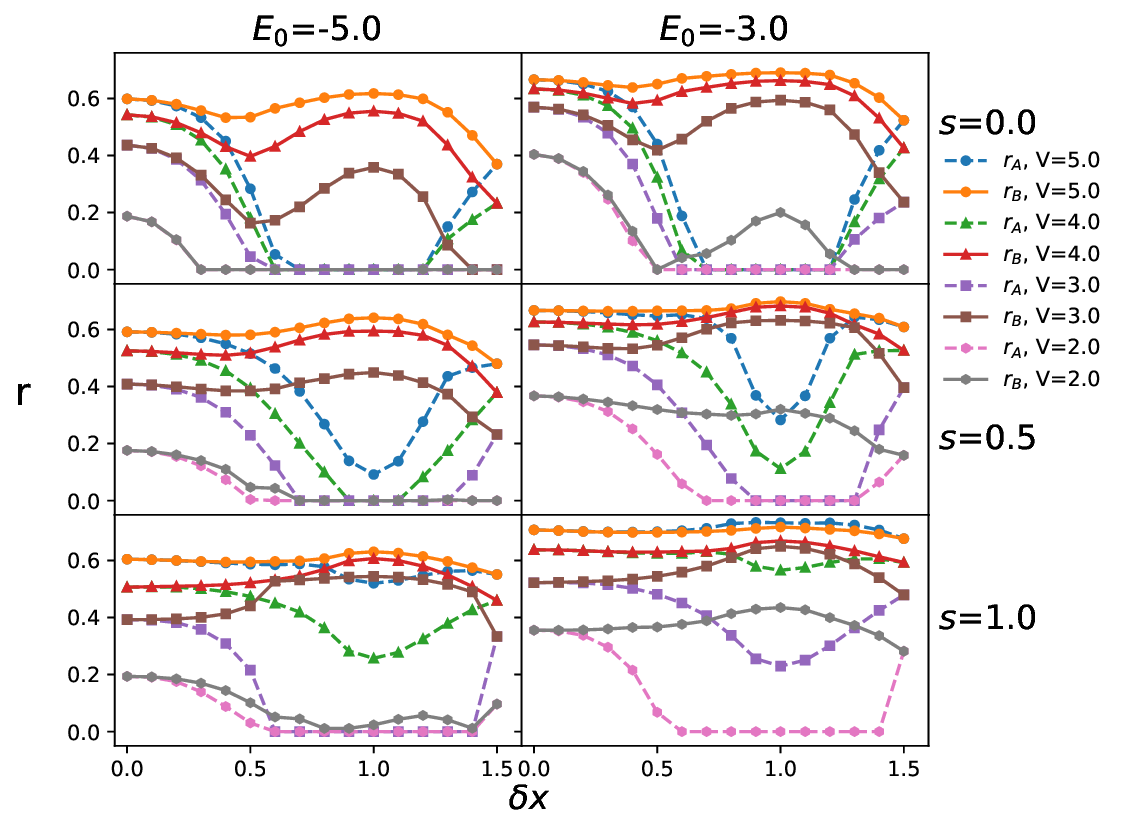}\caption{$\delta x$-dependence of $r_{A}$ and $r_{B}$ for several $V$'s: different decay radius ($d_{s}=(1-s)\sqrt{a_{z}^{2}+a_0^{2}}+s\sqrt{a_{z}^{2}+3a_0^{2}}$):
$s=0$ (NN-cutoff), $s=0.5$, and $s=1$(NNN-cutoff) are used from top to bottom panels, with different on-site $f$-electron energies: $E_{0}=-5$ for the left panels and $E_{0}=-3$ for the right panels.
Note that $a_{z}=1.5a_0$ and the results are very similar to the case of $a_{z}=1.2a_0$.
\label{shift}
}
\end{figure}

\begin{figure}
\centering\includegraphics[scale=0.45]{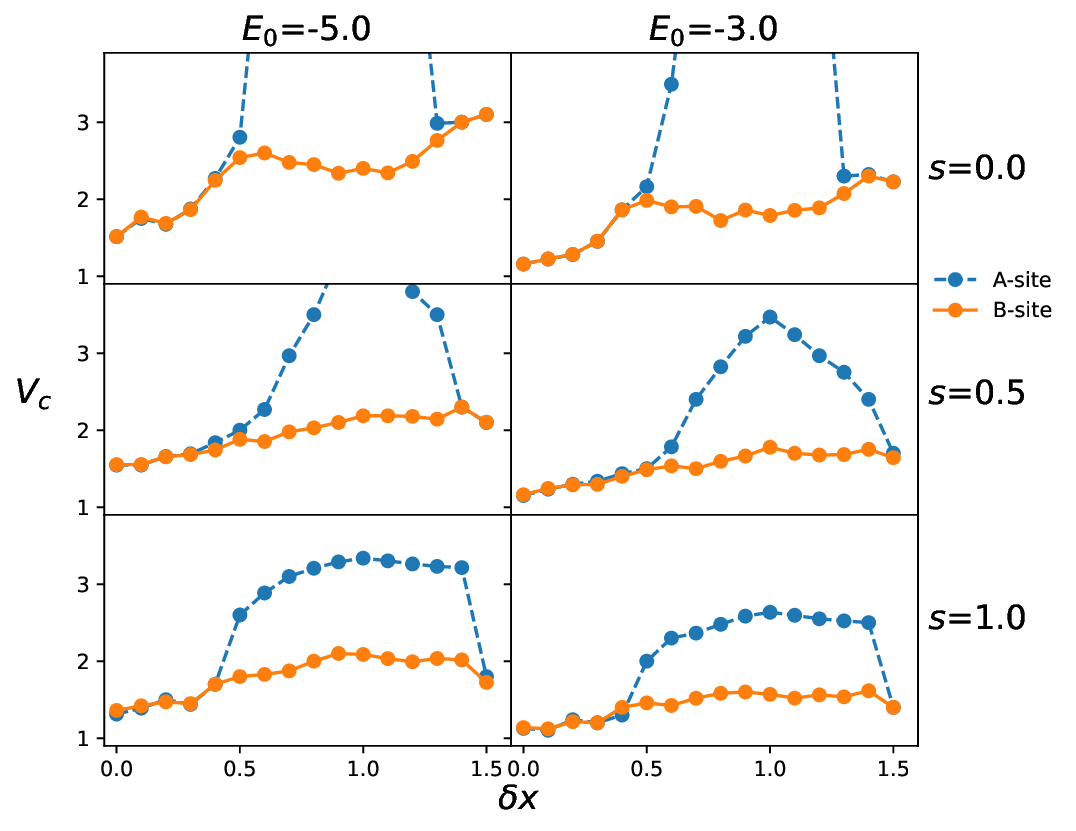}\caption{$\delta x$-dependence critical hybridization strengths: different decay radius ($d_{s}=(1-s)\sqrt{a_{z}^{2}+a_0^{2}}+s\sqrt{a_{z}^{2}+3a_0^{2}}$):
$s=0$ (NN-cutoff), $s=0.5$, and $s=1$(NNN-cutoff) are used from top to bottom, with different on-site $f$-electron energies: $E_{0}=-5$ for the left panels and $E_{0}=-3$ for the right panels. Note that $a_{z}=1.5a_0$.}
\label{critical strengths}
\end{figure}

Here we compare the mean-field solutions obtained by using different cutoff schemes and different $a_z$. In Fig. \ref{shift}, the $\delta x$-dependence of $r_{A}$ and $r_{B}$ for different $V$'s is plotted, using decay radius which interpolate between $\sqrt{a_{z}^{2}+a_0^{2}}$ and $\sqrt{a_{z}^{2}+3a_0^{2}}$, i.e., $d_{\max}\sim d_{s}=(1-s)\sqrt{a_{z}^{2}+a_0^{2}}+s\sqrt{a_{z}^{2}+3a_0^{2}}$, with
$s=0$ (NN-cutoff), $s=0.5$, and $s=1$(NNN-cutoff), from top to bottom panels. We also compare the results obtained by using different on-site $f$-electron energies:
$E_{0}=-5$ for the left panels and $E_{0}=-3$ for the right panels.
Notice that the results obtained by using $a_{z}=1.5a_0$ are very similar to the ones by using $a_{z}=1.2a_0$.
In Fig. \ref{critical strengths}, the $\delta x$-dependence of the critical hybridization strengths $V_{c,A}$ and $V_{c,B}$ is plotted, using  different decay radius
which interpolate between $\sqrt{a_{z}^{2}+a_0^{2}}$ and $\sqrt{a_{z}^{2}+3a_0^{2}}$, i.e., $d_{\max}\sim d_{s}=(1-s)\sqrt{a_{z}^{2}+a_0^{2}}+s\sqrt{a_{z}^{2}+3a_0^{2}}$, with
$s=0$ (NN-cutoff), $s=0.5$, and $s=1$(NNN-cutoff), from top to bottom panels, with  $a_{z}=1.5a_0$ and different on-site $f$-electron energies:
$E_{0}=-5$ for the left panels and $E_{0}=-3$ for the right panels.

\begin{figure}
\centering\includegraphics[scale=0.50]{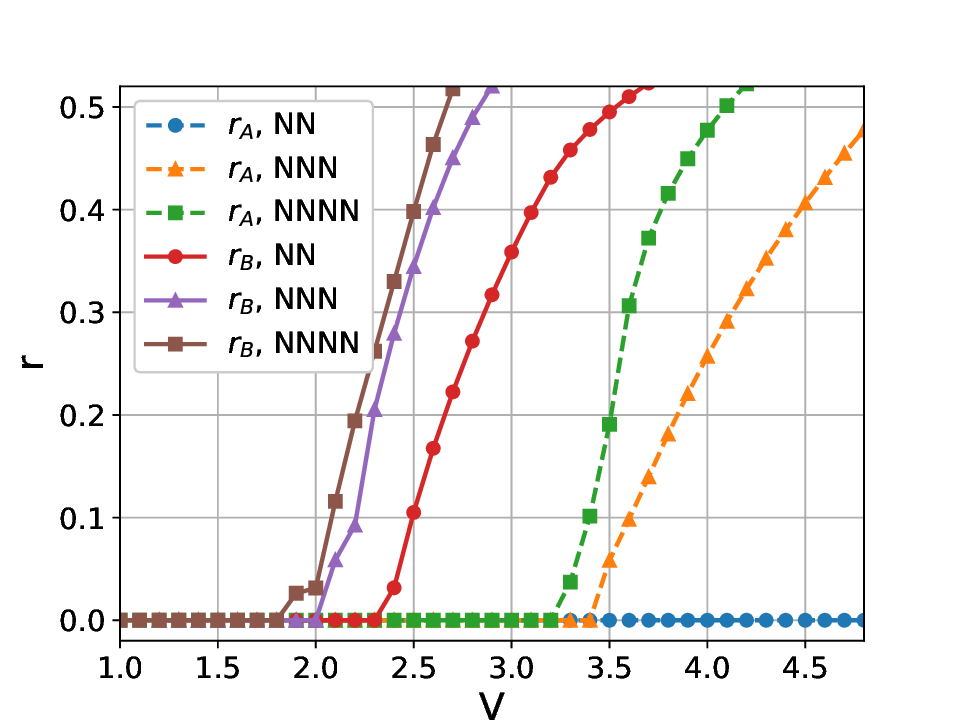}\caption{
{Comparison of the order parameters $r_\eta$ ($\eta = A, B$) versus the nearest-neighbor interlayer hybridization $V_1 = V$ under different cut-off schemes (NN, NNN, and NNNN) in the A-B stack configuration. Here the cutoff radius is $ \xi \sim d_{\max}=\sqrt{a_{z}^{2}+4a_0^{2}}$ within the next to next nearest neighboring (NNNN) sites.
   These results indicate that the critical hybridization strengths are slightly modified by using the different cut-off schemes and the boundaries of the SKS phase remain robust in the presence of long-range interlayer hybridizations. Other parameters are fixed at $a_z = 1.5a_0$, $t_c = 1$, $t_f = 0$, $E_0 = -5$, and $\beta = 400$.}}
\label{NNNN}
\end{figure}

The influence of the long-range hybridizations is further reflected by the critical hybridization strengths in the presence of the next to next nearest neighbor (NNNN) interlayer hybridization as shown in Fig. \ref{NNNN}.  Around the A-B stack configuration, the critical $V_{c,B}$ is slightly reduced from 2.0 to 1.9, while $V_{c,A}$ is reduced from 3.4 to 3.2 (with $E_0=-5$). Hence the boundaries of the SKS phase are only weakly dependent on the cut-off scheme. These results demonstrate the existence of the SKS phase in a wide parameter region and its optimization near the A-B stack configuration although the phase boundaries may deform slightly using different cut-off schemes.

\section{The low-temperature limit of the density of states}

\begin{figure}[ht]
\centering\includegraphics[scale=0.55]{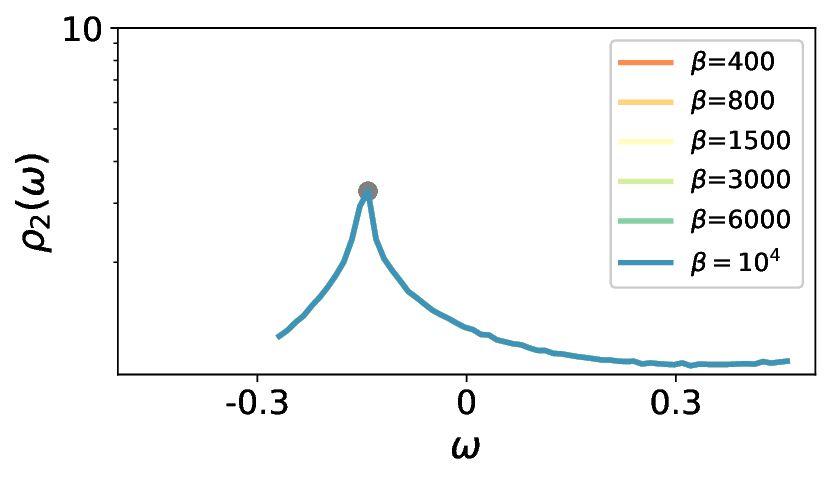}\caption{ A zoom-in view of the DOS of the second band at V=5, with inverse temperatures ranging from $\beta=400$  to $\beta=10^4$. The overlapping curves demonstrate that the DOS remains almost unchanged across different temperature regimes. }
\label{dos_V5}
\end{figure}
\begin{figure}[ht]
\centering\includegraphics[scale=0.55]{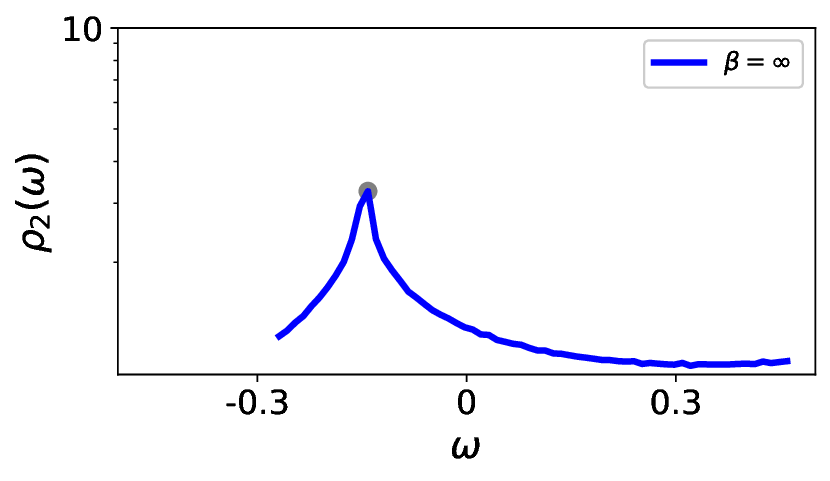}\caption{A zoom-in view of the DOS of the second energy band at V=5, with inverse temperature $\beta=\infty$. The curve represents the ground state electronic structure at the absolute zero temperature. }
\label{dos_V5_2}
\end{figure}

Here we present supplementary data of the density of states (DOS) of the second band in the SKS phase, denoted by $\rho_2(\omega)$.
The temperature dependence of $\rho_2(\omega)$ is implemented in our numerical calculations at a given inverse temperature $\beta$ in the mean-field equations. As shown in Fig. \ref{dos_V5}, our calculations reveal a remarkable temperature independence of $\rho_2(\omega)$ across a wide range of inverse temperatures approaching zero temperature (ranging from $\beta=400$ to $\beta=10^4$). Further analysis at the absolute zero temperature limit ($\beta=\infty$) as shown in Fig. \ref{dos_V5_2} confirms this tendency. It demonstrates the fact that the electronic properties are primarily determined by its intrinsic band structure rather than thermal effects.

\section{Path integral approach and the stability of SKS}

We start from the partition function of the studied model system
\begin{eqnarray}
{\cal Z}={\rm Tr}\{\exp[-\beta (\hat{\cal H}-\mu \hat N)]\}.
\end{eqnarray}
Here, $\hat{\cal H}=\hat{\cal H}[{\hat c},{\hat f}]$ is the system's total Hamiltonian in terms of the original $c$ and $f$ electron (annihilation and creation) operators, $\hat N$ the total particle number operator with $\mu$ being the chemical potential. In the present bilayer honeycomb lattice, each site in the $c$ or $f$ monolayer is labelled by $\textbf i=(\textbf n, \eta)$, with $\textbf n$ labelling the unit cells and $\eta=A,B$ labelling the even or odd sublattices, respectively, so that $\hat{\cal H}$ is essentially four-band model Hamiltonian. In the path integral representation, the inverse temperature $\beta=\frac{1}{k_B T}$ is parameterized by an additional variable $\tau$ introduced as the imaginary time, and the electron operators are represented by the corresponding classical Grassmanian field variables, denoted by $ c_{{\textbf n\eta\sigma}}(\tau)$ and $ f_{{\textbf n\eta\sigma}}(\tau)$, respectively.  The partition function is represented by the functional path integral
\begin{eqnarray}
{\cal Z}=\int {\cal D}[\bar c_A c_A \bar c_B c_B \bar f_A f_A\bar f_B f_B]
e^{-{\cal S}[c_A,c_B,f_A,f_B]},
\end{eqnarray}
where,
${\cal S}[c_A,c_B,f_A,f_B]=\int_0^{\beta}d\tau {\cal L}[c_A,c_B,f_A,f_B]$ is the classical action and
\begin{eqnarray}
{\cal L}[c_A,c_B,f_A,f_B]&=&\sum_{\textbf n \eta \sigma} \bar c_{{\textbf n \eta \sigma}}(\tau)(\frac{\partial}{\partial\tau}-\mu) c_{{\textbf n \eta \sigma}}(\tau)\nonumber\\
&& + \sum_{\textbf n \eta \sigma} \bar f_{{\textbf n \eta \sigma}}(\tau)(\frac{\partial}{\partial\tau}-\mu) f_{{\textbf n \eta \sigma}}(\tau) \nonumber\\
&& + {\cal H}[c_A,c_B,f_A,f_B]
\end{eqnarray}
is the classical Lagrangian, with ${\cal H}[c_A,c_B,f_A,f_B]$ being the classical Hamiltonian obtained by replacing the electron operators by the corresponding electron field variables in the original Hamiltonian, and  ${\cal D}[\bar c_A c_A \bar c_B c_B \bar f_A f_A\bar f_B f_B]$ is the functional integral measure.

In the large $f$ electron Coulomb $U$-limit and employing the slave-boson technique, the $f$ electron operator is decomposed into the auxiliary charged boson operator $\hat b_{\bf{n}\eta} $ and spinful fermion operator $\hat d_{\bf{i}\eta\sigma}$ such that $\hat f_{{\textbf n \eta}\sigma}^{\dag}=\hat d_{\bf{n}\eta\sigma}^{\dag}\hat b_{\bf{n}\eta}$, with the constraint $\hat Q_{\textbf n \eta}\equiv \hat b^{\dag}_{\bf{n}\eta}\hat b_{\bf{n}\eta}+\sum_{\sigma}\hat d^{\dag}_{\bf{n}\eta\sigma}\hat d_{\bf{n}\eta\sigma}=\hat I$ imposed at each lattice sites in the $f$ layer. The constraint is implemented by inserting the following $\delta$-function in the path integral:
\begin{eqnarray}
\delta_{\hat Q_{\textbf n \eta},\hat I}=\frac{1}{2\pi}\int_{-\infty}^{\infty}d\lambda_{\textbf n \eta}e^{i\lambda_{\textbf n \eta}(\hat Q_{\textbf n \eta}-\hat I)}.
\end{eqnarray}
Hence, the partition function is represented by
\begin{eqnarray}
{\cal Z}=\int {\cal D}[\bar c_A c_A \bar c_B c_B \bar d_A d_A\bar d_B d_B b^*_A b_A b^*_B b_B\lambda_A \lambda_B]  \nonumber\\
e^{-{\cal S}_{eff}[c_A,c_B,d_A,d_B,b_A,b_B;\lambda_A,\lambda_B]},
\end{eqnarray}
with the effective action
\begin{eqnarray}
{\cal S}_{eff}[c_A,c_B,d_A,d_B,b_A,b_B;\lambda_A,\lambda_B]\nonumber\\
={\cal S}[c_A,c_B,d_A,d_B,b_A,b_B] \nonumber\\
+\int_0^{\beta}d\tau \sum_{\textbf n \eta}\lambda_{\textbf n \eta}(\tau) [Q_{\textbf n \eta}(\tau)-1],
\end{eqnarray}
and $Q_{\textbf n \eta}(\tau)=b^{*}_{\textbf{n}\eta}(\tau) b_{\textbf{n}\eta}(\tau)+\sum_{\sigma}\bar d_{\textbf{n}\eta\sigma}(\tau)d_{\textbf{n}\eta\sigma}(\tau)$,  ${\cal D}[\lambda_{\textbf n \eta}(\tau)]=\frac{d \lambda_{\textbf n \eta}(\tau)}{2\pi i k_B T}$.

The mean-field approximation employed in the main text corresponds to the semiclassical approximation in the path integral by assuming
$b_{\textbf n \eta}(\tau) =r_{\eta}+\delta b_{\textbf{n}\eta}(\tau)$ and  $\lambda_{\textbf{n},\eta}(\tau)=\lambda_{\eta}+\delta \lambda_{\textbf{n}\eta}(\tau)$, with $(r_{\eta}, \lambda_{\eta})$ being the uniform stationary point of the effective action
\begin{eqnarray}
\frac{\partial}{\partial q_{l}} {\cal S}_{eff}|_{\delta q_{\textbf n l}(\tau)=0}=0.
\end{eqnarray}
In the above, $q_{l}$'s ( with $l=1,2,3,4$) represent the four mean field parameters $(r_A,r_B,\lambda_A,\lambda_B)$, the manifold of the stationary solution, and $\delta q_{\textbf n l}(\tau)=(\delta b_{\textbf n A}(\tau),\delta b_{\textbf n B}(\tau),\delta \lambda_{\textbf n A}(\tau),\delta \lambda_{\textbf n B}(\tau))$ the respective deviations or quantum fluctuations of the field variables from their stationary solution.  For simplicity, the dependence of the effective action on $q_{l}$ will be emphasized as ${\cal S}_{eff}[q]$. Then,
the effective action can be expanded around the stationary point
\begin{eqnarray}
{\cal S}_{eff}[q]={\cal S}^{(0)}_{eff}[q]+\frac{1}{2!}{\cal S}^{(2)}_{eff}[q]+\cdots
\end{eqnarray}
with ${\cal S}^{(0)}_{eff}[q]={\cal S}_{eff}[q]|_{\delta q_{\textbf n l}=0} $ being the stationary action, ${\cal S}^{(2)}_{eff}[q]=\sum_{\textbf n \textbf n' l l'}\int d\tau d\tau'{\cal M}_{\textbf n\textbf n' l l'}
[q](\tau,\tau')\delta q^*_{\textbf n l}(\tau)\delta q_{\textbf n l'}(\tau')$ the correction from the Gaussian-like quantum fluctuation, with
${\cal M}_{\textbf n\textbf n' l l'}[q](\tau,\tau')=\frac{\partial ^2}{\partial \delta q^*_{\textbf n l}(\tau) \partial \delta q_{\textbf n' l'}(\tau')}{\cal S}_{eff}[q]|_{\delta q_{\textbf n l}=0}$.

With these considerations, the partition function can be given by
\begin{eqnarray}
{\cal Z}&=&\int \prod_{\textbf n\eta\sigma}{\cal D}[ \bar c_{\textbf n\eta\sigma} c_{\textbf n\eta\sigma}\bar d_{\textbf n\eta\sigma} d_{\textbf n\eta\sigma}]\prod_{\textbf n l}{\cal D}[\delta q_{\textbf n l}]\\
&& e^{-{\cal S}^{(0)}_{eff}[q]} e^{-\frac{1}{2!}{\cal S}^{(2)}_{eff}[q]+\cdots}. \nonumber
\end{eqnarray}

Neglecting the quantum fluctuations, one obtains the partition function at the mean-field level
\begin{eqnarray}
{\cal Z}^{(0)}=\int \prod_{\textbf n\eta\sigma}{\cal D}[ \bar c_{\textbf n\eta\sigma} c_{\textbf n\eta\sigma} \bar d_{\textbf n\eta\sigma} d_{\textbf n\eta\sigma}]
e^{-{\cal S}^{(0)}_{eff}[q]}.
\end{eqnarray}
Expanding the field variables in the frequency representation and diagonalizing the Hamiltonian matrix in ${\cal S}^{(0)}_{eff}[q]$ in the momentum space using a unitary transformation $\hat U$: $(c_{\textbf k\eta \sigma n}, d_{\textbf k\eta\sigma n})\rightarrow \psi_{\textbf k m\sigma n}$, with $m=1,2,3,4$ being the band index and $\textbf k$ the momentum quantum number valued in the hexagonal BZ, one obtains the stationary action
\begin{eqnarray}
{\cal S}^{(0)}_{eff}[q]=\sum_{mn\sigma} \bar \psi_{\textbf k mn\sigma}[-i\omega_n+E_{\textbf k m\sigma}] \psi_{\textbf k mn\sigma}+\beta E_C,
\end{eqnarray}
where $\omega_n=(2n+1)\pi T$ are the Matsubara frequencies for $n\in \mathbb{Z}$,
$E_{\textbf k m\sigma}$ the eigenvalues depending on the mean field parameters $q_{l}$, as well as for $E_C=L\sum_{\eta}\lambda_{\eta}(r^2_{\eta}-1)$, with $L$ being the total number of unit cells.
Then, we can perform the path integral arriving at
\begin{eqnarray}
{\cal Z}^{(0)}=e^{-\beta E_C}\prod_{\textbf k m\sigma n}\beta[-i\omega_n+E_{\textbf k m\sigma}].
\end{eqnarray}
Taking the logarithm of this expression and then the summation over the Matsubara frequencies we obtain the mean-field free energy $F \equiv -k_B T\ln {\cal Z}^{(0)}$ as
\begin{eqnarray}
F=-\frac{1}{\beta}\sum_{\textbf k m\sigma}\ln [1+e^{-\beta (E_{\textbf k m\sigma}-\mu)}] + E_C~.
\end{eqnarray}

Our numerical solutions of the mean-field parameters $q_{l}=(r_A,r_B,\lambda_A,\lambda_B)$ in the four-band Hamiltonian show the existence of the SKS phase where $r_A=0$ but $r_B>0$ in a wide range of the original model parameters ($V_{c,B}<V<V_{c,A}$). At the mean-field level, such phase is equivalent to the Kondo phase of the reduced three-band Hamiltonian in the A-B stack pattern if the intralayer $f$ electron hopping and the interlayer long-range hybridizations are neglected. In the latter situation, we have
\begin{eqnarray}
\hat{\cal H}[\hat c_A, \hat c_B, \hat f_A,\hat f_B]= \hat{\cal H}[\hat c_A,\hat c_B,\hat f_B]+\hat{\cal H}[\hat f_A],
\end{eqnarray}
implying the complete decoupling of the $A$-sublattice $f$ electrons from the bath. In this case, $\hat{\cal H}[\hat c_A,\hat c_B,\hat f_B]$ is a three-band model without involving the $A$-sublattice $f$ electrons. So the eigenvalue of the decoupled $\hat{\cal H}[\hat f_A]$ part remains at the bare local level $E_0$ well-bellow the Fermi energy without double occupation due to the strong $f$ electron interaction ($U\rightarrow \infty$).

Interestingly, there is a silent discrepancy between the four and three-band model Hamiltonians: in the numerical solutions of the four-band Hamiltonian in the $A$-$B$ stacking pattern, the band energy of the $A$ sublattice $f$ electrons is exactly flat but closes to the Fermi energy when $r_A\rightarrow 0^{+}$, in contrast to the bare level $E_0$ expected from the three-band Hamiltonian.
This discrepancy comes from the mean-field solution of the momentum-independent Lagrange multiplier $\lambda_A$ obtained from the limit $r_A\rightarrow 0^{+}$ in the four-band model as it should keep the effective level $E_0+\lambda_A$ very close to the Fermi level when the hybridization strength $V$ approaches the boundary of the SKS phase.

\begin{figure*}
\centering\includegraphics[scale=0.8]{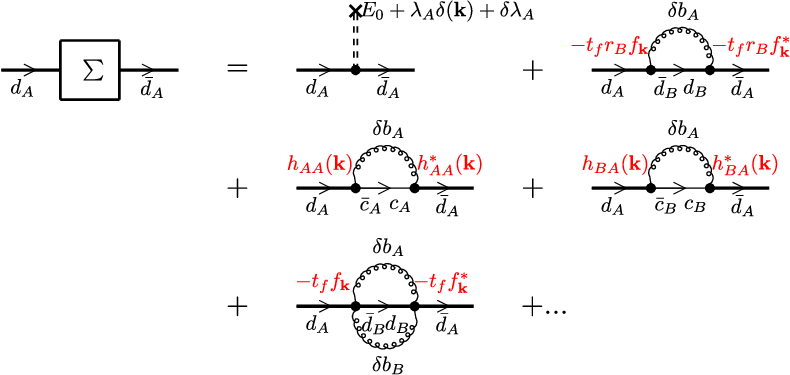}\caption{Self-energy of $f$ electrons in A sublattice. }
\label{self_energy}
\end{figure*}

Here, we argue that for very small $t_f$ and $\alpha$ this issue is linked to the quantum fluctuation of the Lagrangian multiplier field $\lambda_{\textbf k A}(\tau)$ around its mean-field value. To ascertain this point, we consider the fluctuation $\delta \lambda_{\textbf k A}$ while keeping the mean-field solution of $q_{l}$. In this case, by retaining the dependence of the effective action on the field variables, we have
\begin{eqnarray}
&{\cal S}^{(0)}_{eff}[c_A,c_B,d_A,d_B;r_A=0,r_B,\lambda_A,\lambda_B]\\
&={\cal S}^{(0)}_{eff}[c_A,c_B,d_B;r_B,\lambda_B]+{\cal S}^{(0)}_{eff}[d_A;\lambda_A]~\nonumber
\end{eqnarray}
and
\begin{eqnarray}
&{\cal S}^{(2)}_{eff}[c_A,c_B,d_A,d_B,\delta b_A,\delta b_B;r_A=0,r_B,\lambda_A,\lambda_B]\\
&={\cal S}^{(2)}_{eff}[c_A,c_B,d_B,\delta b_B;r_B,\lambda_B]+2{\cal S}^{(2)}_{eff}[\delta b_A;\lambda_A]~.\nonumber
\end{eqnarray}
Here, we only need to consider the diagonal term of the corresponding Hamiltonian in the $f_A$-part:
\begin{eqnarray}
&&{\cal H}^{(0)}_{eff}[d_A;\lambda_A] + \frac{1}{2}{\cal H}^{(2)}_{eff}[\delta b_A;\lambda_A]\nonumber\\
&=&\sum_{\textbf k \sigma }[E_0+(\lambda_A+\delta\lambda_{\textbf k A})]\bar d_{\textbf k A\sigma}(\tau)d_{\textbf k A\sigma}(\tau)\nonumber\\
&+&\sum_{\textbf k}[\delta_{\textbf k A}(\tau)\delta b^*_{\textbf k A}(\tau)\delta b_{\textbf k A}(\tau)-\lambda_A]\nonumber\\
&=&\sum_{\textbf k \sigma}E_0 \bar d_{\textbf k A\sigma}(\tau)d_{\textbf k A\sigma}(\tau)\\
&+&\sum_{\textbf k}\delta \lambda_{\textbf k A}[\sum_{\sigma}\bar d_{\textbf k A\sigma}(\tau)d_{\textbf k A\sigma}(\tau)+\delta b^*_{\textbf k A}(\tau)\delta b_{\textbf k A}(\tau)-1 ]\nonumber
\end{eqnarray}
The last term shows the presence of $\delta\lambda_{\textbf k A}$ just in the place of the mean-field Lagrange parameter $\lambda_A$. Since the integration of the fluctuation $\delta\lambda_{\textbf k A}$ runs over the whole real axis which in turn recovers the $\delta$-function imposed by the no-double occupation constraint of the $A$-sublattice $f$ electrons. The relevant diagrams taking into account of corrections from these fluctuations are shown in Fig. \ref{self_energy} schematically.

\begin{figure*}
\centering\includegraphics[scale=0.8]{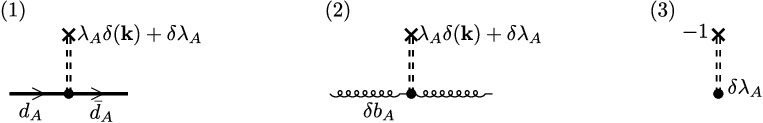}\caption{Contributions to the no-double occupation constraint of the A-sublattice $f$ electrons. }
\label{constraint}
\end{figure*}

Above consideration suggests that in the SKS phase the effective action ${\cal S}^{(0)}_{eff}[q]+\frac{1}{2!} {\cal S}^{(2)}_{eff}[q]$ is similar to the effective action of the three-band interacting Hamiltonian (involving $c_{\textbf k A\sigma}$,$c_{\textbf k B\sigma} $, $d_{\textbf k B\sigma}$ and $r_B, \lambda_B$ ), in addition to the decoupled $A$-sublattice $f$ electron Hamiltonian (involving $d_{\textbf k A\sigma}$, $\delta b_{\textbf k A}$, and $\delta\lambda_{\textbf k A}$). This is due to the linearity of the Lagrangian multiplier appeared in the effective Hamiltonian.  After performing the path integral over $\delta\lambda_{\textbf k A}$, the latter reproduces the representation of the original decoupled Hamiltonian in terms of the $A$-sublattice $f$ electron operator $f_{\textbf k A\sigma}$ with the no-double occupation constraint as illustrated in Fig. \ref{constraint}. Therefore, upon consideration of the quantum fluctuation $\delta \lambda_{\textbf k A}$, the decoupled $f$ electrons remain at the bare level $E_0$ as in the three-band model when $t_f$ and $\alpha$ are sufficiently small. The influence of quantum fluctuations due to sizable $t_f$ and $\alpha$ within or outside the SKS phase ( particularly in the fully decoupled region $0<V<V_{c,B}$) requires further investigation and is beyond the scope of the present study.

\clearpage
\end{appendix}

\bibliography{osks}

\end{document}